\documentclass[aps,pra,preprint,amsmath]{revtex4-1}
\usepackage{silence}
\usepackage{mathrsfs}
\usepackage{graphicx}
\usepackage{epstopdf}
\usepackage{amsthm}
\usepackage{amsmath}
\usepackage{amsfonts}
\usepackage{amssymb}
\usepackage{braket}
\usepackage{bbold}
\usepackage{subfigure}
\usepackage{appendix}
\usepackage{tikz}
\usepackage{bm}
\usepackage{float}
\usepackage{adjustbox}
\def\be{\begin{equation}}
\def\ee{\end{equation}}
\def\ba{\begin{array}}
\def\ea{\end{array}}

\def\1{{\bf{1}}}

\begin{document}
\title{\bf On Sufficient and Necessary Criteria of Multipartite Quantum Entanglement}

\author{Zhi-Bo Chen}
\author{Shao-Ming Fei}

\email{feishm@cnu.edu.cn}
\affiliation{
School of Mathematical Sciences, Capital Normal University, Beijing 100048, China
}

\begin{abstract}
Based on the generalized Bloch representation, we study the separability and entanglement of arbitrary dimensional multipartite quantum states. Some sufficient and some necessary criteria are presented. For certain states, these criteria together are both sufficient and necessary. Detailed examples show that our criteria are better than some existing ones in identifying separability. Based on these criteria, the largest separable ball around the maximally mixed state for arbitrary multi-qubit systems is found, and it is proved that its radius defined via the $l_1$-norm of the Bloch vector is the constant 1. Furthermore, the criteria in this paper can be implemented experimentally.

\end{abstract}

\maketitle

\noindent {\bf 1. Introduction}

Quantum entanglement is a crucial issue in quantum information processing. In particular, multipartite entanglement is an important resource in quantum computation \cite{IE1,IE2,QI}. The detection and quantification of entanglement is one of the key challenges in quantum information science. Although we already have many separability criteria such as PPT \cite{PPT} and measures of entanglement such as concurrence \cite{Conc1,Conc2,ISO}, there are still no both sufficient and necessary criteria and operationally computable entanglement measures for general multipartite systems. Any further progress towards the identification of multipartite entanglement is significant.

In this paper, we investigate the multipartite entanglement for arbitrary dimensional systems. In Section 2, we recall some basic results. In Section 3, we propose sufficient conditions for entanglement \cite{IGE}. In Section 4 and Section 5, we propose necessary conditions for entanglement. These criteria are both sufficient and necessary for some classes of quantum entanglement. In Section 6, we study the separable ball around the maximally mixed state \cite{SB1998,SB2002,SB2003,SB2005,SB2,PRL1998} for $N$-qubit quantum systems. We obtain that the radius of the largest $l_1$ separable ball of multi-qubit systems is $1$. Furthermore, in Section 7, we propose the corresponding entanglement witnesses related to these separability/entanglement criteria and show that these criteria can be implemented experimentally.

\noindent {\bf 2. Preliminary}

Based on the extended Bloch representations of quantum states \cite{KF,EBR,EBR2}, an $n \times n$ density matrix $\varrho$ has the following form,
\begin{align}\label{eq1}
\varrho = \frac{1}{n} \left (I_n + \sum_{i=1}^{n^2 - 1} s_i \lambda_i \right ) ,
\end{align}
where $I_n$ is the $n \times n$ identity matrix and $\{ \lambda_i \}$ are the $n^2 - 1$ generators of $SU(n)$ satisfying $Tr(\lambda_i \lambda_j) = 2\delta_{ij}$. When $\varrho$ is pure state, namely, $ Tr(\varrho^2) = 1$, then $\left| \boldsymbol{s} \right| = \sqrt{\frac{n(n-1)}{2}}$, where $\boldsymbol{s} = (s_1, s_2, ... ,s_{n^2 - 1})$. Set $\widetilde{\lambda_i} = \sqrt{\frac{n(n-1)}{2}} \lambda_i$, then Eq.(1) becomes
$$\varrho = \frac{1}{n} \left ( I_n + \sum_{i=1}^{n^2 - 1} r_i \widetilde{\lambda_i} \right ),~~~  r_i = \sqrt{\frac{2}{n(n-1)}} \ s_i.
$$
$\left| \boldsymbol{r} \right| =|(r_1, r_2, ... ,r_{n^2 - 1})|=1$ for pure states. Notice that when $n>2$, $\left| \boldsymbol{r} \right|=1$ does not necessarily give rise to a density matrix.

Denote $\mathcal{H}_i$ the Hilbert space associated with the $i$th system with dimension $n_i$. An $N$-partite quantum state $\varrho\in\mathcal{H}=\mathcal{H}_1 \otimes \mathcal{H}_2 \otimes ... \otimes\mathcal{H}_N$ with dimension $dim(\mathcal{H}) = n = \prod_{k=1}^{N}n_k$ has the following extended Bloch representation,
\begin{align}\label{eq2}
\varrho = \frac{1}{n}\left ( I_n + \sum_{i_1,i_2...,i_N} \rho_{i_1,i_2,...,i_N} \lambda_{i_1,i_2,...,i_N} \right),
\end{align}
where
\begin{align}
\rho_{i_1,i_2,...,i_N} = n \prod_{k=1}^{N} \frac{1}{Tr  \left ( (\widetilde{\lambda}_{i_k}^{(k)})^2 \right ) } Tr(\varrho  \, \lambda_{i_1,i_2,...,i_N} ),
\end{align}
the summation goes over all $0\leq i_k\leq {n_k}^2-1$, $k=1,2,...,N$, except for $i_1 = i_2 = ... = i_N = 0$, $\lambda_{i_1,i_2,...,i_N} := \widetilde{\lambda}_{i_1}^{(1)} \otimes \widetilde{\lambda}_{i_2}^{(2)} \otimes ... \otimes \widetilde{\lambda}_{i_n}^{(n)}$ with $\widetilde{\lambda}_{0}^{(k)} = I_{n_k}$, $\widetilde{\lambda}_{i_k}^{(k)} = \sqrt{\frac{n_k(n_k -1)}{2}} \lambda_i^{(k)}$, and $\lambda_i^{(k)}$ the $i$th generator of $SU(n_k)$. We call $\rho = (\rho_{i_1,i_2,...,i_N})$ the vector representation of $\varrho$. In this paper, unless otherwise indicated, when referring to a quantum state $\rho$, it is implicitly assumed that $\rho$ is the vector $\rho = (\rho_{0,0,...,1}, \ \rho_{0,0,...,2}  ,...,\rho_{n_1,n_2,...,n_N})$ given by Eq.(3). Denote $\varrho$ the density matrix, $\rho$ is vector representation of $\varrho$. Since $\varrho$ is the matrix of trace 1, the first component $\rho_{0,0,...,0}$ must be 1, which can therefore be omitted unless otherwise indicated. $\mathcal{T} = (\rho_{1,1,...,1}, \ \rho_{1,1,...,2}  ,...,\rho_{n_1,n_2,...,n_N})$ is called correlation tensor \cite{EBR} of $\varrho$.

An $N$-partite density matrix $\varrho$ is fully separable if it can be written as an ensemble of separable states,
$$
\varrho =  \sum_i P_i \varrho_i^{(1)} \otimes \varrho_i^{(2)} \otimes ... \otimes \varrho_i^{(N)},
$$
where $\varrho_i^{(k)}$ is the density matrix associated with the subsystem $k$. For a separable 2-qubit pure state $\varrho = \varrho^A \otimes \varrho^B$, where $\rho^A = (r_1,r_2,r_3) = \boldsymbol{r}$ and $\rho^B = (s_1,s_2,s_3) = \boldsymbol{s}$, we have
$$\rho =  ( \ \underset{\boldsymbol{s}} {\underbrace{\rho_{01},\rho_{02},\rho_{03}}},  \underset{\boldsymbol{r}} {\underbrace{\rho_{10},\rho_{20},\rho_{30}}},  \underset{\boldsymbol{r} \otimes \boldsymbol{s}} {\underbrace{\rho_{11},\rho_{12},...,\rho_{33}}} \ ).$$
Generally we have, see the proof in Appendix A,

{\bf Lemma 1 \ } If an $N$-partite pure state $\rho\in\mathcal{H}=\mathcal{H}_1 \otimes \mathcal{H}_2 \otimes ... \otimes\mathcal{H}_N$ is separable, then $\rho$ has the following form,
$$
\rho = ( \boldsymbol{r}_N , \boldsymbol{r}_{N-1} , ... , \boldsymbol{r}_{1} , \boldsymbol{r}_{1} \otimes \boldsymbol{r}_{2} , \boldsymbol{r}_{1} \otimes \boldsymbol{r}_{3} , ... , \boldsymbol{r}_{N-1} \otimes \boldsymbol{r}_{N} , \boldsymbol{r}_{1} \otimes \boldsymbol{r}_{2} \otimes \boldsymbol{r}_{3} , ... , \boldsymbol{r}_{1} \otimes \boldsymbol{r}_{2} \otimes ... \otimes \boldsymbol{r}_{N} ),
$$
where $ \boldsymbol{r}_{k} = (\rho_{0,...,0,1,0,...0}, \rho_{0,...,0,2,0,...0}, ...,\rho_{0,...,0,{n_k}^2-1,0,...0})$, with all the sub-indices 0 except for the $k$-th index,
$| \boldsymbol{r}_{k} | =1$ for all $1 \leq k \leq N$.

Let $X$ be a set of vectors in the $d$-dimensional real vector space $\mathbb{R}^{d}$. We denote $conv(X) = \{ \sum_{i=1}^{n} P_i\boldsymbol{v}_i \mid \boldsymbol{v}_i \in X,\,n \in \mathbb{N}^+,\, P_i \geq 0,\,\sum_{i=1}^{n}P_i =1\}$ the convex hull of $X$. Let $\mathcal{S}$ be the set of all separable pure states in $\mathcal{H}$, thus $conv (\mathcal{S})$ is the set of all separable states in $\mathcal{H}$. For any $\boldsymbol{a}= (a_i)_{1 \leq i \leq d} \in \mathbb{R}^{d}$, we denote $\left\| \boldsymbol{a} \right\|_p := ( \sum_{i=1}^{d} |a_i|^p )^{\frac{1}{p}}$ the $p$-norm ($1 \leq p$) of $\boldsymbol{a}$. It is straightforward to verify that
\begin{align}
\left\| \boldsymbol{v}_1 \otimes \boldsymbol{v}_2 \otimes... \otimes \boldsymbol{v}_N \right\|_p = \prod_{i=1}^{N} \left\| \boldsymbol{v}_i \right\|_p.
\end{align}
In this paper, the quantum state $\rho$ and the correlation tensor $\mathcal{T}$ are always considered as real vectors given by Eq.(3). Their $p$-norms given by
$$ \left\| \rho \right\|_p =  \left ( \sum_{i,j,...,k} | \rho_{i,j,...,k} |^p \right )^{\frac{1}{p}},  \; \; \left\| \mathcal{T} \right\|_p = \left ( \sum_{i,j,...,k \neq 0} | \rho_{i,j,...,k} |^p \right )^{\frac{1}{p}},
$$
where the sum does not include the first component $\rho_{0,0,...,0} = 1$.

\noindent {\bf 3. Necessary Conditions for Separability}

In this section we give sufficient criteria for entanglement. We have, see the proof in Appendix B,

{\bf Theorem 1 \ }
If an $N$-partite quantum state $\rho \in \bigotimes_k  \mathcal{H}_k $ is fully separable, then correlation tensor $\mathcal{T}$ of $\rho$ satisfies the following inequalities,
\begin{align}
\begin{cases} \left\| \mathcal{T} \right\|_p \leq \prod_{k=1}^{N}{m_k}^{\frac{1}{p}-\frac{1}{2}}, & \text{if} \ 1 \leq p \leq 2, \\ \left\|
 \mathcal{T} \right\|_p \leq 1, & \text{if} \ p>2,  \end{cases}
\end{align}
where $m_k = {n_k}^2 -1$. In particular, for $p=1$ and $p=2$ we have
\begin{align}
\left\| \mathcal{T} \right\|_1 \leq \prod_{k=1}^{N} \sqrt{{n_{k}}^2 -1},~~~
\left\| \mathcal{T} \right\|_2 \leq 1 \ .
\end{align}

Define
$$
t_{i,j,...,k} = Sgn(\rho_{i,j,...,k}) := \begin{cases} 1, & \text{if} \ \rho_{i,j,...,k}>0, \\ 0, & \text{if} \ \rho_{i,j,...,k}=0, \\ -1, & \text{if} \ \rho_{i,j,...,k}<0. \end{cases}
$$
We have the following more detailed theorem, see the proof in Appendix C.

{\bf Theorem 2 \ }
If an $N$-partite quantum state $\rho \in \bigotimes_k  \mathcal{H}_k $ is fully separable, then the correlation tensor $\mathcal{T}$ of $\rho$ satisfies the following inequality,
\begin{align}
 \left\| \mathcal{T} \right\|_1 \leq M(\rho) \ ,
\end{align}
where
\begin{align} \notag
&M(\rho)=\frac{1}{N}max \left\{ \sum_{i_2,i_3,...,i_N} |t_{1,i_2,i_3,...,i_N}| \ , \sum_{i_2,i_3,...,i_N} |t_{2,i_2,i_3,...,i_N}| \ ,..., \sum_{i_2,i_3,...,i_N} |t_{{n_1,i_2,i_3,...,i_N}}| \right\} \\
& + \frac{1}{N}max \left\{ \sum_{i_1,i_3,...,i_N} |t_{i_1,1,i_3,...,i_N}| \ , \sum_{i_1,i_3,...,i_N} |t_{i_1,2,i_3,...,i_N}| \ ,..., \sum_{i_1,i_3,...,i_N} |t_{{i_1,n_2,i_3,...,i_N}}| \right\} + ...... \notag \\
& + \frac{1}{N}max \left\{ \sum_{i_1,i_2,...,i_{N-1}} |t_{i_1,i_2,...,i_{N-1},1}| \ , \sum_{i_1,i_2,...,i_{N-1}} |t_{i_1,i_2,...,i_{N-1},2}| \ ,..., \sum_{i_1,i_2,...,i_{N-1}} |t_{{i_1,i_2,...,i_{N-1},n_N}}| \right\} . \notag
\end{align}

We illustrate Theorem 2 with two examples.

{\bf Example 1 } The $d \times d$ isotropic states is a one-parameter family of  bipartite quantum states and can be written as $\varrho_{iso}^{\alpha} = \frac{1-\alpha}{d^2}I_{d^2} + \alpha \ket{\phi^+} \bra{\phi^+}$, where $\ket{\phi^+} = \frac{1}{\sqrt{d}} \sum_{i=1}^d \ket{ii}$ and $\alpha \in (0,1)$. For the state $\varrho_{iso}^{\alpha}$, one has $\rho_{i,j} = 0$ for $i\neq j$, and $\rho_{i,i} = \pm \alpha \frac{1}{d-1}$ for $1 \leq i \leq d^2-1$ \cite{KF}. Then we have $M\left ( \rho_{iso}^{\alpha} \right ) = 1$ and $\left\| \mathcal{T} \right\|_1 = \alpha \frac{d^2-1}{d-1} = (d+1) \alpha $. According to Theorem 2, when $\alpha > \frac{1}{d+1}, \ \varrho_{iso,d}^{\alpha}$ is entangled, which is consistent with the fact that all $d \times d$ isotropic states are separable if and only if $\alpha \leq \frac{1}{d+1}$ \cite{ISO}.

{\bf Example 2 } Consider 3-qubit state $\varrho_A^\alpha = \alpha A + (1-\alpha)\frac{I_8}{8}$, where $\alpha \in \left( 0 , \frac{1}{\sqrt{2}} \right]$ and $A = \frac{1}{8} \left ( I_8  + \sigma_1 \otimes \sigma_3 \otimes \sigma_1 + \sigma_2 \otimes \sigma_2 \otimes \sigma_2\right )$. We have $\left\| \mathcal{T} \right\|_1 = 2\alpha$ and $M(\rho_A^\alpha) = 1$. By Theorem 2, when $\frac{1}{2} < \alpha \leq \frac{1}{\sqrt{2}}$, $\varrho_A^\alpha$ is entangled.
In fact, combining the results from Theorem 3 below, we have that $\varrho_A^\alpha$ is entangled if and only if $\frac{1}{2} < \alpha \leq \frac{1}{\sqrt{2}}$.
Observation 4 in \cite{NJP} perfectly detects the entanglement of the noisy GHZ state, but fails to detect the entanglement of $\varrho_A^\alpha$.

For most quantum states, Theorem 2 is stronger than Theorem 1. Nevertheless, Theorem 1 has the advantage of having a fixed upper bound, which will be discussed in detail in Section 7.

\noindent {\bf 4. Sufficient Conditions for Separability of Multi-qubit Systems}

In Section 3 we discussed general quantum states without dimensional restriction. In this part, however, we will first discuss $N$-qubit systems and then generalize to multipartite systems in general. The proofs of the theorems in this section are based on the conclusions from group representation theory, and in Appendices D and E, we give the relation between separable states and characters of finite group. For multi-qubit systems, we have the following sufficient conditions for separability, see the proof in Appendix F.

{\bf Theorem 3 \ }
For an $N$-qubit state $\rho$, if $\left\| \rho \right\|_1 \leq 1$, then $\rho$ is fully separable.

{\bf Example 3}
Consider the 3-qubit state $\varrho_A^\alpha$ given in Example 2, where it has been shown that $\varrho_A^\alpha$ is entangled when $\frac{1}{2} < \alpha \leq \frac{1}{\sqrt{2}}$ by Theorem 2. By Theorem 3, we have that $\varrho_A^\alpha$ is fully separable when $ \alpha \leq \frac{1}{2}$. These two criteria from Theorems 2 and 3 together are both sufficient and necessary for this state. Consider $\varrho_A^\alpha$ as a $\mathbb{C}^2 \otimes \mathbb{C}^4$ state. One has that $\varrho_A^\alpha$ is bipartite separable when $\alpha \leq \frac{1}{\sqrt{12}} \approx 0.2886$ by the Corollary 2 in Ref.\cite{NS} and also by the Proposition 3 in Ref.\cite{KF}; when $\alpha \leq \frac{1}{\sqrt{14}} \approx 0.2672$ by the separable ball criterion \cite{SB2002}, and $\varrho_A^\alpha$ is fully separable when $\alpha \leq \frac{1}{\sqrt{22}} \approx 0.2132$ by Corollary 3 in \cite{SB2005}. Thus, our Theorem 3 detects entanglement better than the above criteria, see Figure 1.
\begin{figure}[!hbtp]
    \centering
    \includegraphics[width=1\linewidth]{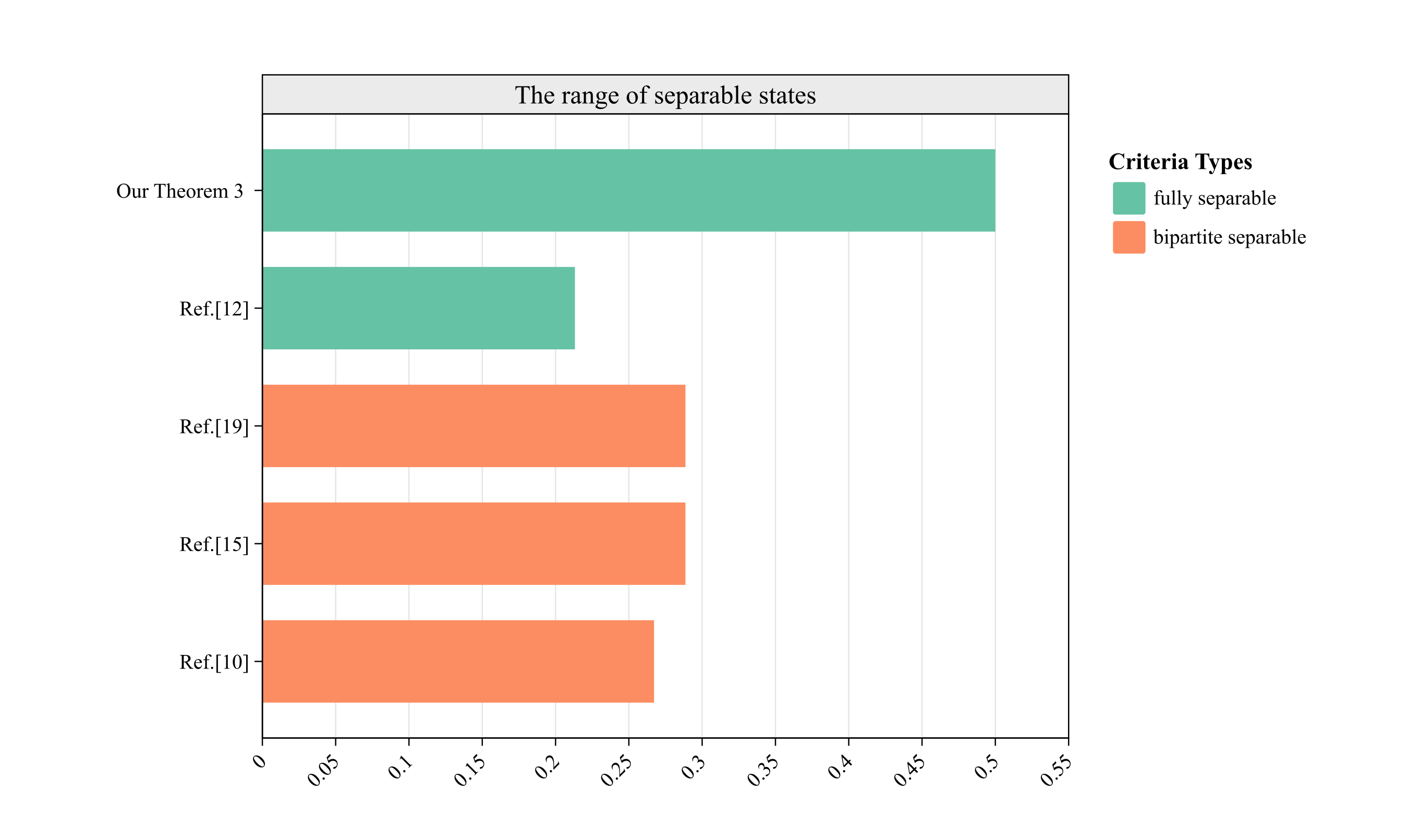}
    \caption{ The horizontal coordinates indicate the range of parameter $\alpha$ can detect the separability of $\varrho_A^\alpha$ in Example 3.
    }
    \label{fig:enter-label}
\end{figure}

In particular, for 3-qubit and 4-qubit states we have the following stronger sufficient conditions on separability. The proofs of Theorem 4 and Theorem 5 are given in Appendix G.

{\bf Theorem 4 \ }
For any 3-qubit state $\rho$, if there exist $i,j,k \in \{ 1,2,3 \}$ such that $Sgn(\rho_{0,j,k}) \cdot Sgn(\rho_{i,0,k}) \cdot Sgn(\rho_{i,j,0}) \geq 0$, then $\rho$ is fully separable if
$$
\left\| \rho \right\|_1 \leq 1 + 2min\left\{ |\rho_{0,j,k}|, |\rho_{i,0,k}|, |\rho_{i,j,0}| \right\}.
$$

{\bf Theorem 5 \ }
If there exist distinct $i_1,i_2,i_3 \in \{ 1,2,3 \}$, distinct $j_1,j_2,j_3 \in \{ 1,2,3 \}$ and distinct $k_1,k_2,k_3 \in \{ 1,2,3 \}$ such that $Sgn(\rho_{0,j_t,k_t}) \cdot Sgn(\rho_{i_t,0,k_t}) \cdot Sgn(\rho_{i_t,j_t,0}) \geq 0$ for all $1 \leq t \leq 3$, then a 3-qubit state $\rho$ is fully separable if
\begin{align} \notag
\left\| \rho \right\|_1 \leq &
1 + 2min\left\{ |\rho_{0,j_1,k_1}|, |\rho_{i_1,0,k_1}|, |\rho_{i_1,j_1,0}| \right\} + 2min\left\{ |\rho_{0,j_2,k_2}|, |\rho_{i_2,0,k_2}|, |\rho_{i_2,j_2,0}| \right\} \\
& + 2min\left\{ |\rho_{0,j_3,k_3}|, |\rho_{i_3,0,k_3}|, |\rho_{i_3,j_3,0}| \right\} . \notag
\end{align}

Analogous to Theorem 4, we have the following conclusion for 4-qubit states, see the proof in Appendix H.

{\bf Theorem 6 \ }
For any 4-qubit state $\rho$, if there exist $i,j,k,l \in \{ 1,2,3 \}$ such that
$$
\begin{array}{l}
Sgn(\rho_{i,j,0,0}) Sgn(\rho_{i,0,k,0}) Sgn(\rho_{i,0,0,l}) Sgn(\rho_{0,j,k,0}) Sgn(\rho_{0,j,0,l}) Sgn(\rho_{0,0,k,l}) Sgn(\rho_{i,j,k,l}) \geq 0,\\
Sgn(\rho_{i,j,0,0}) Sgn(\rho_{i,0,k,0}) Sgn(\rho_{0,j,k,0}) \geq 0, \\
Sgn(\rho_{i,j,0,0}) Sgn(\rho_{i,0,0,l}) Sgn(\rho_{0,j,0,l}) \geq 0,\\
Sgn(\rho_{i,0,k,0})  Sgn(\rho_{i,0,0,l}) Sgn(\rho_{0,0,k,l}) \geq 0,
\end{array}
$$
then $\rho$ is fully separable if
$$
\left\| \rho \right\|_1 \leq 1 + 2 \cdot lessmid \left\{| \rho_{i,j,0,0} |, \ | \rho_{i,0,k,0} |, \ | \rho_{i,0,0,l} |, \ | \rho_{0,j,k,0}|, \ | \rho_{0,j,0,l}|, \ | \rho_{0,0,k,l}|, \ | \rho_{i,j,k,l}| \right\},
$$
where $lessmid$ represents the summation of the numbers in the set that are less than the median, namely, the summation of the smallest, the second smallest and the third smallest numbers.

Theorem 5 is stronger than Theorem 4 for states that the conditions in Theorem 5 are satisfied. Similar to Theorem 5 for 3-qubit case, a stronger theorem than Theorem 6 can also be given for the 4-qubit case under the same conditions. Theorem 4, Theorem 5 and Theorem 6 are strictly stronger than Theorem 3 for the 3-qubit and 4-qubit cases.

{\bf Example 4}  Consider the 3-qubit noisy GHZ state, $\varrho_{GHZ}^{\alpha} = \frac{1-\alpha}{8}I + \alpha \ket{GHZ} \bra{GHZ}$, where $\ket{GHZ} = \frac{1}{\sqrt{2}}  \left ( \ket{000} + \ket{111} \right )$ and $\alpha \in (0,1)$. By Eq.(3), the vector representation of $\varrho_{GHZ}^{\alpha}$ is
$\rho_{033} = \alpha$, $\rho_{303} = \alpha$, $\rho_{330}= \alpha$, $\rho_{221} = \rho_{212} = \rho_{122} = -\alpha$, $\rho_{111} = \alpha$ and $\mbox{other} \ \rho_{ijk} = 0$.
Let $i=j=k=3$, from Theorem 4 we have that $\varrho_{GHZ}^{\alpha}$ is fully separable when $\alpha \leq \frac{1}{5}$, which is consistent with the best result for the state \cite{GHZS}. Our result is better than $\alpha \leq 0.1139$ given in \cite{SB2005}, and $\varrho_{GHZ}^{\alpha}$ is bipartite separable when $\alpha \leq \frac{1}{7}$ \cite{SB2002}.

{\bf Example 5} Consider the 4-qubit noisy GHZ state, $\varrho_{GHZ_4}^{\alpha} = \frac{1-\alpha}{16}I + \alpha \ket{GHZ} \bra{GHZ}$, where $\ket{GHZ} = \frac{1}{\sqrt{2}}  \left ( \ket{0000} + \ket{1111} \right )$. From Theorem 6 we have that $\varrho_{GHZ_4}^{\alpha}$ is separable when $\alpha \leq \frac{1}{9}$, which is in consistent with the best best result presented in \cite{GHZS}. Our result is better than $\alpha \leq 0.0436$ in \cite{SB2005}, and $\varrho_{GHZ_4}^{\alpha}$ is bipartite separable when $\alpha \leq \frac{1}{15}$ \cite{SB2002}.

{\bf Example 6}
Now consider the $2 \otimes 4$ bound entangled state given in Ref.\cite{S}, $\varrho_{a,\alpha} = \alpha \varrho_A + \frac{1-\alpha}{8}I_8$, where $\varrho_A = \frac{7a}{7a+1} \varrho_{ent} + \frac{1}{7a+1} \ket{\phi} \bra{\phi}$, $a \in \left [ 0,1 \right ]$, with $\ket{\phi} = \ket{1} \otimes \left ( \sqrt{\frac{1+a}{2}} \ket{0} + \sqrt{\frac{1-a}{2}} \ket{2} \right )$ and $\varrho_{ent} = \frac{2}{7}\sum_{i=1}^3 \ket{\psi_i} \bra{\psi_i} + \frac{1}{7} \ket{03} \bra{03}$, $\ket{\psi_i} = \frac{1}{\sqrt{2}} \left ( \ket{0} \ket{i-1} + \ket{1} \ket{i}  \right )$, $i=1,2,3$. Let us consider the state $\varrho_{a,\alpha}$ as a 3-qubit state. Using the Pauli matrix as basis, from Theorem 4 we have that $\varrho_{a,\alpha}$ is fully separable for all $a$  when $\alpha \leq \frac{1}{5}$. This result is better than that of Corollary 2 in Ref.\cite{SB2002}, which fails to detect the separability when $a < \frac{1}{224} \left ( \sqrt{3745} - 49 \right )$; and Corollary 1 in Ref.\cite{S} which can only detect bipartite separability for $\alpha \leq \frac{1}{5}$, see detailed calculations in Appendix L.

In Example 4 and Example 5, it is verified that Theorem 4 and Theorem 6 together are sufficient and necessary separability criteria for the noisy GHZ states. In fact, Theorem 4, Theorem 5 and Theorem 6 can be also generalized to $N$-qubit systems, giving rise to sufficient and necessary separability criterion for $N$-qubit noisy GHZ states, see Appendix J. Such generalization is not unique, more multipartite separability criteria can be obtained through character table, see Appendix D.

\noindent {\bf 5. Sufficient Conditions for Separability of Arbitrary Quantum states}

Next we generalize the above theorems to arbitrary dimensional multipartite systems.
To ensure that $\varrho$ is an $N$-partite (semi-positive) state in $\mathcal{H}=\mathcal{H}_1 \otimes \mathcal{H}_2 \otimes ... \otimes\mathcal{H}_N$, we use a different Bloch representation. Let $\check{\lambda}_0^{(k)}= I_{n_k}$, $\check{\lambda}_i^{(k)}= \sqrt{\frac{n_k}{2(n_k - 1)}} \lambda_i^{(k)}$ and $\check{\lambda}_{i_1,i_2,...,i_N}= \check{\lambda}_{i_1}^{(1)} \otimes \check{\lambda}_{i_2}^{(2)} \otimes ... \otimes \check{\lambda}_{i_N}^{(N)}$, then
 \begin{align}\label{eq10}
\varrho = \frac{1}{n}\left ( I_n + \sum_{i_1,i_2...,i_N} \rho_{i_1,i_2,...,i_N} \cdot \check{\lambda}_{i_1,i_2,...,i_N} \right).
\end{align}
We denote $\sideset{_s}{}{\mathop{\rho}} = (\rho_{i_1,i_2,...,i_N})$ the vector representation of $\varrho$ to distinguish it from the previous vector representation.

It is shown that $\boldsymbol{r}$ is a vector representation of a density matrix if $ \left\| \boldsymbol{r} \right\|_2 \leq \sqrt{\frac{n}{2(n-1)}}$ in Ref.\cite{KF}, i.e., $\left\| \sideset{_s}{}{\mathop{\boldsymbol{r}}} \right\|_2 \leq 1 $. Hence, it is still possible to define simple separable states $\rho_1 \otimes \rho_2 \otimes ... \otimes \rho_N$, where $\sideset{_s}{}{\mathop{\rho_i}} \in \left\{ ( \pm 1,0,...,0), (0, \pm 1,0,...,0), ... , \ (0,...,0, \pm 1) \right\}$, see Appendix D. The only difference is that at this point the simple separable states are no longer pure states, but this does not cause any difficulty. Thus, by using the vector representation given in Eq.(8), Theorem 3, Theorem 4, and Theorem 5 can all be generalized to arbitrary multipartite quantum systems, with proofs similar to those given in Appendices E, F and G.

{\bf Theorem 7 \ } For arbitrary quantum state $\varrho\in\mathcal{H}=\mathcal{H}_1 \otimes \mathcal{H}_2 \otimes ... \otimes\mathcal{H}_N$, we have

(1) If $\left\| \sideset{_s}{}{\mathop{\rho}} \right\|_1 \leq 1$, then $\rho$ is fully separable.

(2) When $N=3$, if there exist $i \in \{ 1,2,...,{n_1}^2 -1 \}, \ j \in \{ 1,2,...,{n_2}^2 -1 \}, \ k \in \{ 1,2,...,{n_3}^2 -1 \},\ $ such that $Sgn(\rho_{0,j,k}) \cdot Sgn(\rho_{i,0,k}) \cdot Sgn(\rho_{i,j,0}) \geq 0$, then $\rho$ is fully separable if
$$\left\| \sideset{_s}{}{\mathop{\rho}} \right\|_1 \leq 1 + 2min\left\{ |\rho_{0,j,k}|, |\rho_{i,0,k}|, |\rho_{i,j,0}| \right\} . $$

(3) When $N=3$, if there exist distinct $i_1,i_2,i_3 \in \{ 1,2,...,{n_1}^2 -1 \}$, distinct $j_1,j_2,j_3 \in \{ 1,2,...,{n_2}^2 -1 \}$ and distinct $k_1,k_2,k_3 \in \{ 1,2,...,{n_3}^2 -1 \}$, such that $Sgn(\rho_{0,j_t,k_t}) \cdot Sgn(\rho_{i_t,0,k_t}) \cdot Sgn(\rho_{i_t,j_t,0}) \geq 0$ for all $1 \leq t \leq 3$, then $\rho$ is fully separable if
\begin{align} \notag
\left\| \sideset{_s}{}{\mathop{\rho}} \right\|_1 \leq &
1 + 2min\left\{ |\rho_{0,j_1,k_1}|, |\rho_{i_1,0,k_1}|, |\rho_{i_1,j_1,0}| \right\} + 2min\left\{ |\rho_{0,j_2,k_2}|, |\rho_{i_2,0,k_2}|, |\rho_{i_2,j_2,0}| \right\} \\
& + 2min\left\{ |\rho_{0,j_3,k_3}|, |\rho_{i_3,0,k_3}|, |\rho_{i_3,j_3,0}| \right\} . \notag
\end{align}

Theorem 6 can also be generalized to $N \geq 4$ and will not be repeated here. The criteria presented in Section 4 and Section 5 give not only sufficient conditions for separability, but also the detailed separable pure state decompositions, as given in the proofs in Appendices E, F, G and H.

\noindent {\bf 6. Separable Balls}

A fundamental question in the study of quantum entanglement is to ascertain the geometry of the set of all separable states. For $N$-qubit systems, based on the extended Bloch representation, we are able to show that the set of all separable pure states $\mathcal{S} \subseteq \mathbb{R}^{4^N-1}$ is a bounded 2$N$-dimensional real manifold as well as a regular 2$N$-dimensional real affine algebraic variety. However, it is still difficult to calculate the set of all separable states $conv(\mathcal{S})$.

A key aspect of the geometry of a convex set is the size of the largest ball that fits inside the convex set. The first separable ball was found in \cite{SB1998}, which was subsequently proved in \cite{SB2002} to be the largest separable ball for bipartite quantum systems. Specifically, for $\mathbb{C}^{d_1} \otimes \mathbb{C}^{d_2}$ systems, if $Tr(\varrho^2) \leq \frac{1}{d_1 d_2 -1}$, then $\varrho$ is a separable state. Roughly speaking, a separable ball is a $p$-norm ball around the maximally mixed state, $B_a:= \left\{ \left\| \varrho - \frac{I}{n} \right\|_p \leq a \mid \varrho \ \mbox{is density matirx} \right\}$, such that $B_a \subseteq conv(\mathcal{S})$. The authors in Ref.\cite{SB1998,SB2002,SB2003,SB2005,SB2,PRL1998} have studied separable balls for bipartite and multipartite systems. To date, no exact radius has been found yet for the largest separable ball around the maximally mixed state in general for arbitrary dimensional multipartite systems.

The separable ball given in \cite{SB2002,SB2003} involves the computation of eigenvalues, which becomes difficult for high dimensions. We define a more reasonable $l_p$ separable ball in the Bloch representation space, which does not involve eigenvalue computation. Denote $V_{\mathcal{H}}$ the Bloch vector representation space of a given quantum system $\mathcal{H}$, which is a real vector space spanned by the basis in Eq.(2) except for the identity matrix $I$. An $l_p$ separable ball is $B_p(a):= \left\{  \rho \in V_{\mathcal{H}} \mid \left\| \rho \right\|_p \leq a \right\}$ such that $B_p(a) \subseteq conv(\mathcal{S})$, i.e., all $\rho$ are separable states, where $ \left\| \cdot \right\|_p$ is the $p$-norm. For $N$-qubit systems, this definition is equivalent to the one given in \cite{SB2002,SB2003} when $p=2$, i.e., the well-known separable ball defined by purity \cite{SB1998}. In fact, this is true for any multipartite quantum system by changing the normalization of the basis, see Eq.(18) in Appendix I. Specifically, for $N$-qubit system one has $ Tr(\varrho ^2) = \frac{1}{2^N} \left ( 1+ {\left\| \rho \right\|_2}^2 \right ) $, where $\rho$ is vector representation of density matrix $\varrho$. Then the separable ball criterion \cite{SB2002} becomes, if $\left\| \rho \right\|_2 \leq \sqrt{\frac{1}{2^N -1}}$, then $\rho$ is bipartite separable for under any bipartitions.

Denote $R(l_p)$ the radius of the largest $l_p$ separable ball. Using the criteria in this paper, we have the following elegant conclusion.

{\bf Theorem 8 \ }
For $N$-qubit systems, in terms of the extended Bloch representation given in Eq.(3), $B_1(1) = \left\{  \rho \in V_{\mathcal{H}} \mid \left\| \rho \right\|_1 \leq 1 \right\}$ is the largest separable ball, namely, $R(l_1)=1$.

{\it Proof \ } By Theorem 3, $B_1(1)$ is indeed a separable ball. We prove that it is the largest one. Consider an $N$-qubit state $\varrho = \frac{1}{2^N}  \left (  I + {\sigma_1}^{\otimes N} + \epsilon \left ( {\sigma_2}^{\otimes N}  \right ) \right )$ with $\epsilon > 0$. The vector representation of $\varrho$ is $\rho = (0,...,0,1,0,...,0,\epsilon,0,...,0)$. We have $\left\| \rho \right\|_1 = \left\| \mathcal{T} \right\|_1 = 1+\epsilon$, where $\mathcal{T}$ is correlation tensor. By Theorem 2, we have that $\varrho$ is entangled when $\epsilon > 0$. Thus, $B_1(1)$ is indeed the largest $l_1$ separable ball. $\Box$

Next, we estimate the range of $R(l_p)$ for $N$-qubit systems. These conclusions can be easily generalized to arbitrary high-dimensional multipartite systems with more complicated notations. By Refs.\cite{SB2002,SB2003,SB2005}, we have
$$
\sqrt{\frac{1}{\frac{4}{9} \cdot 3^{N} -1}} \leq R(l_2) \leq \sqrt{\frac{1}{2^N -1}},
$$
and $ R(l_\infty ) \leq ( \frac{1}{\sqrt{3}} )^N $ by Eq.(6). Obviously, every separable ball gives a sufficient condition for separability, as shown in Theorem 3. It is seen that the $l_1$ separable ball criterion (Theorem 3) is stronger than the $l_2$ (purity) separable ball criterion \cite{SB1998} in many cases, as shown in Example 4, Example 5 and Example 6. For $N$-qubit systems, if the vector representation $\rho$ satisfies $| \rho_{i,j,...,k} | \geq \frac{1}{2^N -1}$ for all $i,j,...,k$, or $\rho$ has at most $2^N-1$ nonzero components, then $\left\| \rho \right\|_2 \leq \sqrt{\frac{1}{2^N -1}}$ implies $\left\| \rho \right\|_1 \leq 1$. Since Theorem 4, Theorem 5 and Theorem 6 are stronger than Theorem 3 for the corresponding cases, these criteria are stronger than the criterion in \cite{SB1998} for most cases. Unlike other separable balls \cite{PRL1998,SB2003} whose radius decreases exponentially, the radius of $l_1$ separable ball does not decrease with $N$, but is the constant 1. This ensures that the criterion remains effective as the number of partite increases, which characterizes certain geometry structure of the space containing all separable quantum states.

Corresponding to the largest separable ball, we define the minimal ball containing all separable states. If $B_p(a) := \left\{  \rho \in V_{\mathcal{H}} \mid \left\| \rho \right\|_p \leq a \right\}$ satisfies $conv(\mathcal{S}) \subseteq B_p(a) $, then $B_p(a)$ is an $l_p$ ball containing all separable states. Denote $R_E(l_p)$ the radius of the $l_p$ ball containing all separable states with the smallest radius.

{\bf Theorem 9 \ }
For $N$-qubit systems, we have
$$R_E(l_p) = \begin{cases} \left ( \left( 3^{1 - \frac{p}{2}}  +1 \right )^N - 1 \right )^{\frac{1}{p}}, & \text{if} \ 1 \leq p \leq 2, \\ \left ( 2^N - 1\right )^{\frac{1}{p}} , & \text{if} \ p > 2. \end{cases}$$

The proof is given in Appendix K. We conjecture that every $l_p$ ball in Theorem 9 is the smallest $l_p$ ball that contains all $N$-qubit states, see discussions in Appendix K.

It is seen that as the number of subsystems increases, $R(l_1)$ remains constant but $R_E(l_1)$ increases; $R(l_2)$ decreases but $R_E(l_2)$ increases. Since $\mbox{the largest separable ball} \ \subseteq conv(\mathcal{S}) \subseteq \ \mbox{minimal ball containing all separable states} $, as the dimension increases $conv(\mathcal{S})$ becomes increasingly complex and difficult to describe. This is also in consistent with the fact that the entanglement identification is only perfectly known for 2-qubit systems.

\noindent {\bf 7. Experimental Realization of Criteria Based on 1-norm}

In this section, we show that any $l_1$-norm-based criterion can be realized by an hermitian matrix (witness) $W$, making the criterion experimentally implementable. Denote $V_{\mathcal{H}}$ the hermitian matrix space, with the vector representations given by a set of basis. Then $\left\| \rho \right\|_1 = f_{\rho}(\rho)$ for any given state $\rho$, where $f_{\rho} = \sum_i Sgn(\rho_i)\rho_i $, and $f_\rho \in {V_{\mathcal{H}}}^{*}$, where ${V_{\mathcal{H}}}^{*}$ is the dual space of $V_{\mathcal{H}}$. Thus, by the Riesz representation theorem, we have $\left\| \rho \right\|_1 = f_{\rho}(\rho) = Tr(\varrho W_{\rho})$. The following theorem gives a concrete construction of $W_{\rho}$ in terms of the extended Bloch representation.

{\bf Theorem 10 \ }
For any state $\varrho\in\mathcal{H}=\mathcal{H}_1 \otimes \mathcal{H}_2 \otimes ... \otimes\mathcal{H}_N$ given in Eq.(2), with the vector representation $\rho$, set
$$
W = -aI_n + \sum_{i_1,...,i_N} \frac{n \cdot Sgn(\rho_{i_1,...,i_N})}{Tr \left ( {\lambda_{i_1,...,i_N}}^2 \right ) } \lambda_{i_1,...,i_N} \ ,
$$
then we have the relation $\left\| \rho \right\|_1 - a = Tr(\varrho W)$, where $a$ is an arbitrary real number, $n = \prod_k dim \left ( \mathcal{H}_k \right ) $.

{\it Proof \ }
Since $\left\{ \lambda_{i_1,...,i_N} \right\}$ is an orthogonal basis, from Eq.(3) it is straightforward to verify.  $\Box$

In particular, for $N$-qubit systems, $W = -aI + \sum_{i_1,...,i_N} \left ( \pm \sigma_{i_1} \otimes \sigma_{i_2} \otimes ... \otimes \sigma_{i_N} \right ) $. For any given state $\rho$, our criterion gives an entanglement(or separability) witness $W_{\rho}$. Next, we illustrate the specific construction of $a$ in Theorem 10. For example, if one uses Theorem 3 to detect separability, the coefficient $a$ is constant $1$. To use Theorem 1 and Theorem 2 to detect entanglement, one needs to slightly modify $W$ in Theorem 10:
$$
W_\rho = -aI_n + \sum_{i_1,...,i_N \neq 0} \frac{n \cdot Sgn(\rho_{i_1,...,i_N})}{Tr \left ( {\lambda_{i_1,...,i_N}}^2 \right ) } \lambda_{i_1,...,i_N} \ ,
$$
similarly, we have the relation $\left\| \mathcal{T} \right\|_1 - a = Tr(\varrho W)$. If one uses Theorem 1 or Theorem 2 to detect entanglement, the coefficient $a$ is $\prod_{k=1}^{N} \sqrt{{dim(\mathcal{H}_k)}^2 -1}$ in Eq.(6) or $M(\rho)$ in Eq.(7), respectively. By Theorem 10 all criteria based on $1$-norm in this paper can be formulated by witnesses: if $Tr(\varrho W_{\rho}) > 0$ then $\varrho$ is entangled; if $Tr(\varrho W'_{\rho}) \leq 0$ then $\varrho$ is separable. Here, $W_{\rho}$ is explicitly given by $\rho$.

Concerning $\rho_{i_1,...,i_N}$, we may randomize the plus and minus signs to get a set of measurements: $\left\{ W = -aI_n + \sum_{i_1,...,i_N} \pm \frac{n}{Tr \left ( {\lambda_{i_1,...,i_N}}^2 \right ) } \lambda_{i_1,...,i_N}  \right\}$. If there exists a $W$ such that $Tr(\varrho W) > 0$, we obtain $\left\| \rho \right\|_1 - a > 0$, namely, we can use the entanglement criterion based on $1$-norm.

\noindent {\bf 8. Conclusion and Discussion}

We have proposed criteria for identifying multipartite entanglement. They are both sufficient and necessary for the entanglement detection in certain classes of quantum states. Moreover, it has been shown that the radius of the largest $l_1$ separable ball is $1$ for $N$-qubit systems, where the radius is defined via the $1$-norm of the Bloch vector representation.

The criteria in Section 4 are all based on the convex hull of a set of separable states, that is, finding a subset of $conv\left ( \mathcal{S} \right )$ to detect the separability. Obviously, finding more separable states leads to a larger convex hull, i.e., a stronger separability criterion. For $N$-qubit systems, the convex hull derived from Theorem 3, i.e., the separable ball in Theorem 8, is $4^N-1$-dimensional, which means that the (real) dimension of $conv\left ( \mathcal{S} \right )$ is $4^N-1$. However, it remains unclear whether $conv\left ( \mathcal{S} \right )$ is a $4^N-1$-dimensional real manifold. Exactly the same, by Theorem 7 for general quantum systems $\mathcal{H}$ the dimension of $conv\left ( \mathcal{S} \right )$ is $n^2 - 1$. Thus, no proper linear subspace of the density matrix space contains all the separable pure states.

Notice that unlike $2$-norm and Ky-Fan norm which are local unitary invariants \cite{KF}, $1$-norm is not a local unitary invariant, which allows us to use optimization so we can get better results than in examples of this paper. This is an advantage that the Ky-Fan norm-based criterion such as \cite{PPT, KF, EBR, Extended1, Extended2, Extended3} does not have.

The investigation on detection of multipartite quantum entanglement has attracted much attention, with potential applications in quantum information processing and quantum computation. Our results may highlight further studies on entanglement identification, as well as on the entanglement structures and distributions in multipartite systems.

\newpage
\noindent {\bf Appendix A}

{\bf Proof of Lemma 1}

If $\rho$ is separable pure state, then $\rho = \rho^{(1)} \otimes \rho^{(2)} \otimes ... \otimes \rho^{(N)}$, where $ \rho^{(k)} \in \mathcal{H}_k$ is pure state. Under basis $ \{ \frac{1}{n_k} I,   \frac{1}{n_k} \widetilde{\lambda}_{1}^{(k)} , \frac{1}{n_k} \widetilde{\lambda}_{2}^{(k)}, ... , \frac{1}{n_k} \widetilde{\lambda}_{{n_k}^2 -1}^{(k)} \}$, the coordinate of $ \rho^{(k)} $ is $ \rho^{(k)} = (1,\boldsymbol{r}_k)$, and $ | \boldsymbol{r}_{k} | =1$. The basis in Eq.(2) is generated by those basis:
$$ \bigotimes_{k=1}^{N} \left\{ \frac{1}{n_k} I,   \frac{1}{n_k} \widetilde{\lambda}_{1}^{(k)} , \frac{1}{n_k} \widetilde{\lambda}_{2}^{(k)}, ... , \frac{1}{n_k} \widetilde{\lambda}_{{n_k}^2 -1}^{(k)} \right\} ,$$
so the coordinate of $\rho$ under this basis is $(1,\boldsymbol{r}_1) \otimes (1,\boldsymbol{r}_2) \otimes ... \otimes (1,\boldsymbol{r}_N)$.
Adjusting the order of components, vector representation of $\rho$ can be written as:
$$\rho = ( \boldsymbol{r}_N , \boldsymbol{r}_{N-1} , ... , \boldsymbol{r}_{1} , \boldsymbol{r}_{1} \otimes \boldsymbol{r}_{2} , \boldsymbol{r}_{1} \otimes \boldsymbol{r}_{3} , ... , \boldsymbol{r}_{N-1} \otimes \boldsymbol{r}_{N} , \boldsymbol{r}_{1} \otimes \boldsymbol{r}_{2} \otimes \boldsymbol{r}_{3} , ... , \boldsymbol{r}_{1} \otimes \boldsymbol{r}_{2} \otimes ... \otimes \boldsymbol{r}_{N} ) ,$$
specifically, $\boldsymbol{r}_{k} \otimes \boldsymbol{r}_{l}$ is $(\rho_{0,...,0,i,0,...,0,j,0,...,0})_{1 \leq i \leq {n_k}^2-1, \ 1 \leq j \leq {n_l}^2-1 }$, where $i,j$ are at the $k$th, $l$th positions respectively; $\boldsymbol{r}_{k} \otimes \boldsymbol{r}_{l} \otimes \boldsymbol{r}_{m}$ is$(\rho_{0,...,0,i,0,...,0,j,0,...,0,t,0,...,0})_{1 \leq i \leq {n_k}^2-1, \ 1 \leq j \leq {n_l}^2-1, \ 1 \leq t \leq {n_m}^2-1 }$ where $i,j,t$ are at the $k$th, $l$th, $m$th positions respectively. The same applies to the rest. $\Box$ \\

\noindent {\bf Appendix B}

{\bf Proof of Theorem 1}

{\bf Lemma S1 \ }
If an $N$-partite pure state $\rho \in \bigotimes_k  \mathcal{H}_k $ is fully separable, $dim(\mathcal{H}_k) = n_k$, $\mathcal{T}$ is correlation tensor of $\rho$, then the following inequalities hold:
$$
\begin{cases} \left\| \mathcal{T} \right\|_p \leq \prod_{k=1}^{N} \left ( {n_k}^2 - 1 \right )^{\frac{1}{p}-\frac{1}{2}}, & \text{if} \ 1 \leq p \leq 2, \\ \left\| \mathcal{T} \right\|_p \leq 1, & \text{if} \ p>2.  \end{cases}
$$

{\it Proof \ }
By Lemma 1, $\mathcal{T} = \bigotimes_{k=1}^N  \boldsymbol{r}_k$. Let $m_k := {n_k}^2 - 1$, now $\boldsymbol{r}_k \in \mathbb{R}^{m_k}$, and $\left\| \boldsymbol{r}_k \right\|_2 = 1 $, we have
$$ \begin{cases} \left\| \boldsymbol{r}_k \right\|_p \leq {m_k}^{\frac{1}{p}-\frac{1}{2}}, & \text{if} \ 1 \leq p \leq 2, \\ \left\| \boldsymbol{r}_k \right\|_p \leq 1, & \text{if} \ p>2,  \end{cases} $$
when $1 \leq p \leq 2$ the maximum value is obtained at $\boldsymbol{r}_k = (\frac{1}{\sqrt{m_k}}, \frac{1}{\sqrt{m_k}},...,\frac{1}{\sqrt{m_k}})$; when $p>2$ the maximum value is obtained at $\boldsymbol{r}_k = (1, 0, 0,...,0)$. By Eq.(4), $\left\| \mathcal{T} \right\|_p = \prod_{k=1}^N \left\| \boldsymbol{r}_k  \right\|_p$, so the lemma is proved.  $\Box$

Now if $\rho \in \bigotimes_k  \mathcal{H}_k $ is fully separable, then $\rho = \sum_i P_i \rho_i$, thus $\mathcal{T} = \sum_i P_i \mathcal{T}_i$, where $\mathcal{T}_i$ is correlation tensor of pure state $\rho_i$ . By the convexity of $p$-norm, we have $\left\| \mathcal{T} \right\|_p =  \left\| \sum_i P_i \mathcal{T}_i \right\|_p \leq \sum_i P_i \left\| \mathcal{T}_i \right\|_p $, Theorem 1 is proved by Lemma S1.  $\Box$  \\

\noindent {\bf Appendix C}

{\bf Proof of Theorem 2}

We first define a function. For any $N$-partite state $\rho \in \bigotimes_{k=1}^N \mathcal{H}_k$, $\mathcal{T}$ is correlation tensor of $\rho$, $m_k := (dim(\mathcal{H}_k))^2 -1$. $f_{\mathcal{T}} (\boldsymbol{x}) := \sum_{i,j,...,k} t_{i,j,...,k} \cdot x_{i,j,...,k} $, where $\boldsymbol{x} \in \bigotimes_k \mathbb{R}^{m_k}$ are real tensors, $x_{i,j,...,k}$ are components of tensor $\boldsymbol{x}$, $t_{i,j,...,k} = Sgn(\rho_{i,j,...,k}) $ is depends on $\rho$. Notice that $Sgn(\rho_{i,j,...,k}) \cdot \rho_{i,j,...,k} = |\rho_{i,j,...,k}| $, then $f_\mathcal{T}(\mathcal{T}) = \left\| \mathcal{T} \right\|_1$. Function $f_\mathcal{T}$ may vary for different $\mathcal{T}$. $f_\mathcal{T}$ is always linear homogeneous polynomial function, thus it is obviously convex function. $f_\mathcal{T}$ can provide a criterion. First, consider a simple 3-qubit case. For any separable pure state $\rho$, $\boldsymbol{s} = \boldsymbol{x} \otimes \boldsymbol{y} \otimes \boldsymbol{z}$ is correlation tensor of $\rho$, where $\boldsymbol{x}=(x_1,x_2,x_3), \boldsymbol{y}=(y_1,y_2,y_3), \boldsymbol{z}=(z_1,z_2,z_3), \ \left\| \boldsymbol{x} \right\|_2 = \left\| \boldsymbol{y} \right\|_2 = \left\| \boldsymbol{z} \right\|_2 = 1$, then
$$ f_\mathcal{T}(\boldsymbol{s}) = \sum_{1 \leq i,j,k \leq 3}t_{ijk} \cdot x_i y_j z_k  \ \leq \sum_{1 \leq i,j,k \leq 3}|t_{ijk}| \ |x_i| \ |y_j| \ |z_k| , $$
by mean inequality and $|x_i|, |y_j|, |z_k| \leq 1 $ for all $1 \leq i,j,k \leq 3$, we have
$$
f_\mathcal{T}(\boldsymbol{s}) \leq  \sum_{1 \leq i,j,k \leq 3}|t_{ijk}| \cdot \frac{1}{3} (|x_i|^3 + |y_j|^3 + |z_k|^3)  \leq \sum_{1 \leq i,j,k \leq 3} |t_{ijk}| \cdot \frac{1}{3} (|x_i|^2 + |y_j|^2 + |z_k|^2) \ , $$
then we have
\begin{align} \notag
f_\mathcal{T}(\boldsymbol{s}) & \leq  \left ( \frac{|x_1|^2}{3} \sum_{1 \leq j,k \leq 3} |t_{1jk}| \right ) + \left ( \frac{|x_2|^2}{3} \sum_{1 \leq j,k \leq 3} |t_{2jk}| \right ) + \left ( \frac{|x_3|^2}{3} \sum_{1 \leq j,k \leq 3} |t_{3jk}| \right ) + \notag \\
& \; \; \; \; \left ( \frac{|y_1|^2}{3} \sum_{1 \leq i,k \leq 3} |t_{i1k}| \right ) + ... +\left ( \frac{|z_3|^2}{3} \sum_{1 \leq i,j \leq 3} |t_{ij3}| \right ) \ , \notag
\end{align}
and
\begin{align} \notag
& \left ( \frac{|x_1|^2}{3} \sum_{1 \leq j,k \leq 3} |t_{1jk}| \right ) + \left ( \frac{|x_2|^2}{3} \sum_{1 \leq j,k \leq 3} |t_{2jk}| \right ) + \left ( \frac{|x_3|^2}{3} \sum_{1 \leq j,k \leq 3} |t_{3jk}| \right )  \\
& \leq \frac{1}{3}max \left\{ \sum_{1 \leq j,k \leq 3} |t_{1jk}| \ , \sum_{1 \leq j,k \leq 3} |t_{2jk}| \ , \sum_{1 \leq j,k \leq 3} |t_{3jk}| \right\} \cdot \left ( |x_1|^2+|x_2|^2+|x_3|^2 \right ) \ , \notag
\end{align}
and  $\left\| \boldsymbol{x} \right\|_2 = \left\| \boldsymbol{y} \right\|_2 = \left\| \boldsymbol{z} \right\|_2 = 1$, thus, we finally conclude that
\begin{align} \notag
f_\mathcal{T}(\boldsymbol{s}) \leq & \frac{1}{3}max \left\{ \sum_{j,k} |t_{1jk}| \ , \sum_{j,k} |t_{2jk}| \ , \sum_{j,k} |t_{3jk}| \right\}
+ \frac{1}{3}max \left\{ \sum_{i,k} |t_{i1k}| \ , \sum_{i,k} |t_{i2k}| \ , \sum_{i,k} |t_{i3k}| \right\} \\
& + \frac{1}{3}max \left\{ \sum_{i,j} |t_{ij1}| \ , \sum_{i,j} |t_{ij2}| \ , \sum_{i,j} |t_{ij3}| \right\} \ . \notag
\end{align}
If $\rho$ is separable, $\rho = \sum_i P_i \rho_i$, then $\mathcal{T} =  \sum_i P_i \boldsymbol{s}_i$, where $\mathcal{T}$ is correlation tensor of $\rho$, $\boldsymbol{s}_i$ is correlation tensor of separable pure state $\rho_i$. And $\left\| \mathcal{T} \right\|_1 = f_\mathcal{T}(\mathcal{T}) = f_\mathcal{T}(\sum P_i \boldsymbol{s}_i) = \sum P_i f_\mathcal{T}(\boldsymbol{s}_i)$, therefore we have:
\begin{align} \notag
\left\| \mathcal{T} \right\|_1 \leq & \frac{1}{3}max \left\{ \sum_{j,k} |t_{1jk}| \ , \sum_{j,k} |t_{2jk}| \ , \sum_{j,k} |t_{3jk}| \right\}
+ \frac{1}{3}max \left\{ \sum_{i,k} |t_{i1k}| \ , \sum_{i,k} |t_{i2k}| \ , \sum_{i,k} |t_{i3k}| \right\} \\
& + \frac{1}{3}max \left\{ \sum_{i,j} |t_{ij1}| \ , \sum_{i,j} |t_{ij2}| \ , \sum_{i,j} |t_{ij3}| \right\} \notag ,
\end{align}
this conclusion can be generalized to arbitrary quantum state, and the proof is not much different from the above.

For arbitrary separable pure state, by Lemma 1, its correlation tensor is $ \boldsymbol{s} = \boldsymbol{s}^{(1)} \otimes \boldsymbol{s}^{(2)} \otimes ... \otimes \boldsymbol{s}^{(N)} = \left ( s_{i_1}^{(1)} s_{i_2}^{(2)}... \ s_{i_N}^{(N)} \right )_{i_1,i_2,...,i_N}$, where $\boldsymbol{s}^{(k)} = (s_1^{(k)}, s_2^{(k)}, ..., s_{n_k}^{(k)})$, $ \left\| \boldsymbol{s}^{(k)} \right\|_2 = 1$ for all $k$.
$$ f_\mathcal{T}(\boldsymbol{s}) = \sum_{i_1,i_2,...,i_N} t_{i_1,i_2,...,i_N} \cdot  s_{i_1}^{(1)} s_{i_2}^{(2)}... \ s_{i_N}^{(N)}  \ \leq \sum_{i_1,i_2,...,i_N}|t_{i_1,i_2,...,i_N}| \ |s_{i_1}^{(1)}| \ |s_{i_2}^{(2)}| ... |s_{i_N}^{(N)}| ,$$
by mean inequality and $|s_{i_k}^{(k)}| \leq 1 $ for all $1 \leq i_k \leq {n_k}^2-1, \ 1 \leq k \leq N$, we have
\begin{align} \notag
f_\mathcal{T}(\boldsymbol{s}) & \leq  \sum_{i_1,i_2,...,i_N}|t_{i_1,i_2,...,i_N}| \cdot \frac{1}{N} \left (  |s_{i_1}^{(1)}|^N + |s_{i_2}^{(2)}|^N + ... + |s_{i_N}^{(N)}|^N  \right )  \notag \\
& \leq \sum_{i_1,i_2,...,i_N} |t_{i_1,i_2,...,i_N}|  \frac{1}{N} \left (  |s_{i_1}^{(1)}|^2 + |s_{i_2}^{(2)}|^2 + ... + |s_{i_N}^{(N)}|^2  \right ) \notag \\
& \leq  \left ( \frac{|s_1^{(1)}|^2}{N} \sum_{i_2,i_3,...,i_{N}} |t_{1,i_2,...,i_{N}}| \right ) + \left ( \frac{|s_2^{(1)}|^2}{N} \sum_{i_2,i_3,...,i_{N}} |t_{2,i_2,...,i_{N}}| \right ) + ...... \notag \\
& \; \; \; \; \;  + \left ( \frac{|s_{n_1}^{(1)}|^2}{N} \sum_{i_2,i_3,...,i_{N}} |t_{n_1,i_2,...,i_{N}}| \right ) + \left ( \frac{|s_1^{(2)}|^2}{N} \sum_{i_1,i_3,...,i_{N}} |t_{i_1,1,i_3,...,i_{N}}| \right ) + ...... \notag \\
& \; \; \; \; \; + \left ( \frac{|s_{n_N}^{(N)}|^2}{N} \sum_{i_1,i_2,...,i_{N-1}} |t_{i_1,...,i_{N-1},n_N}| \right ) \notag \\
& \leq \frac{ \sum_u |s_{u}^{(1)}|^2  }{N}max \left\{ \sum_{i_2,...,i_{N}} |t_{1,i_2,...,i_{N}}| \ , \sum_{i_2,...,i_{N}} |t_{2,i_2,...,i_{N}}| \ ,..., \sum_{i_2,...,i_{N}} |t_{n_1,i_2,...,i_{N}}| \right\} \notag \\
& \; \; \; \; \; +......+ \notag \\
&\frac{ \sum_u |s_{u}^{(N)}|^2  }{N}max \left\{ \sum_{i_1,...,i_{N-1}} |t_{i_1,...,i_{N-1},1}| \ , \sum_{i_1,...,i_{N-1}} |t_{i_1,...,i_{N-1},2}| \ ,..., \sum_{i_1,...,i_{N-1}} |t_{i_1,...,i_{N-1},n_N}| \right\}  \notag
\end{align}
and  $\left\| \boldsymbol{s}^{(1)} \right\|_2 = \left\| \boldsymbol{s}^{(2)} \right\|_2 = ... = \left\| \boldsymbol{s}^{(N)} \right\|_2 = 1$, thus, we finally conclude that

\begin{align} \notag
f_\mathcal{T}(\boldsymbol{s}) \leq &  \frac{1}{N}max \left\{ \sum_{i_2,i_3,...,i_N} |t_{1,i_2,i_3,...,i_N}| \ , \sum_{i_2,i_3,...,i_N} |t_{2,i_2,i_3,...,i_N}| \ ,..., \sum_{i_2,i_3,...,i_N} |t_{{n_1,i_2,i_3,...,i_N}}| \right\} \\
& + \frac{1}{N}max \left\{ \sum_{i_1,i_3,...,i_N} |t_{i_1,1,i_3,...,i_N}| \ , \sum_{i_1,i_3,...,i_N} |t_{i_1,2,i_3,...,i_N}| \ ,..., \sum_{i_1,i_3,...,i_N} |t_{{i_1,n_2,i_3,...,i_N}}| \right\} \notag \\
& + ...... \notag \\
& + \frac{1}{N}max \left\{ \sum_{i_1,...,i_{N-1}} |t_{i_1,...,i_{N-1},1}| \ , \sum_{i_1,...,i_{N-1}} |t_{i_1,...,i_{N-1},2}| \ ,..., \sum_{i_1,...,i_{N-1}} |t_{{i_1,...,i_{N-1},n_N}}| \right\} . \notag
\end{align}
If $\rho$ is separable, $\rho = \sum_i P_i \rho_i$, then $\mathcal{T} =  \sum_i P_i \boldsymbol{s}_i$, where $\mathcal{T}$ is correlation tensor of $\rho$, $\boldsymbol{s}_i$ is correlation tensor of separable pure state $\rho_i$. And $f_\mathcal{T}(\sum P_i \boldsymbol{s}_i) = \sum P_i f_\mathcal{T}(\boldsymbol{s}_i)$, therefore we have: $
\left\| \mathcal{T} \right\|_1 = f_\mathcal{T}(\mathcal{T}) = \sum P_i f_\mathcal{T}(\boldsymbol{s}_i) \leq M(\rho) . \; \; \Box  $  \\

\noindent {\bf Appendix D}

{\bf Definition \ }
$\left\{ \rho_i \right\}$ are pure qubit states, $\rho_i$ is vector representation a density matrix, if all $\rho_i$ satisfy
$$\rho_i \in \left\{ (1,0,0), (-1,0,0), \ (0,1,0), \ (0,-1,0), \ (0,0,1), \ (0,0,-1) \right\} ,$$
then $\rho_1 \otimes \rho_2 \otimes ... \otimes \rho_N$ is called simple separable state.

Simple separable state has a simplified expression where the zero components are ignored. For example, without omitting the first component 1, $\rho_1 = (1,x,0,0), \ \rho_2 = (1,0,y,0)$, then $\rho_1 \otimes \rho_2 = (1,0,y,0,x,0,0,0,xy,0,0,0,0,0,0,0)$ is simple separable state, where $x,y= \pm1$. (Recall Lemma 1, reordered here.) Consider projection $\pi: \rho_1 \otimes \rho_2 \mapsto (1,x,y,xy)$, $(1,x,y,xy)$ is called reduced expression of $\rho_1 \otimes \rho_2$. In general, for $N$-qubit, by ignoring the zero components, simple separable state have reduced expression of order $N$:
$$\left ( 1,x^{(N)},x^{(N-1)},...,x^{(1)},x^{(1)} x^{(2)},x^{(1)} x^{(3)},...,x^{(N-1)} x^{(N)},x^{(1)} x^{(2)} x^{(3)},...,x^{(1)} x^{(2)}...x^{(N)} \right ) ,$$
each simple separable state corresponds to a unique reduced expression, while a single reduced expression can correspond to $3^N$ different simple separable states. Thus, by determining all possible reduced expressions, we can actually determine all possible simple separable states. This is achievable. In the case of 2-qubit,
\begin{equation}
    \bordermatrix{%
	       & 1       & x       &y      &xy    \cr
	s_1   & 1       & 1       & 1     & 1    \cr
	s_2   & 1       & 1       & -1    & -1   \cr
	s_3   & 1       & -1      & 1     & -1   \cr
	s_4   & 1       & -1      & -1    & 1
    }
\end{equation}
above row vectors $s_1, s_2, s_3, s_4$ are all possible reduced expression of 2-qubit simple separable state. The rows and columns of the matrix being orthogonal, it is a general phenomenon, in fact it is a character table of a finite abelian group.

{\bf Definition \ }
Simple separable state group of level $N$ is a finite abelian group
$$G = \left< x^{(1)}, x^{(2)}, ... , x^{(N)} \ \big| \ {x^{(i)}}^2 = 1, \ x^{(i)} x^{(j)} = x^{(j)} x^{(i)} \ \mbox{for all} \ 1 \leq i,j \leq N \right> $$
where $G$ is a group generated by the generators $\left\{ x^{(1)}, x^{(2)}, ... , x^{(N)} \right\}$ and the equivalence relations are $ {x^{(i)}}^2 = 1, \ x^{(i)} x^{(j)} = x^{(j)} x^{(i)}$ for all $1 \leq i,j \leq N$. $G$ is a finite abelian group, and
$G =  \left\{ 1,x^{(N)},x^{(N-1)},...,x^{(1)},x^{(1)} x^{(2)},x^{(1)} x^{(3)},...,x^{(N-1)} x^{(N)},x^{(1)} x^{(2)} x^{(3)},...,x^{(1)} x^{(2)}...x^{(N)} \right\}$.

{\bf Lemma S2 \ }
Considering character of group representation \cite{GRO} as vector, there is a one-to-one correspondence between reduced expressions of simple separable state and irreducible character of simple separable state group: \\
$\left\{\mbox{reduced expressions of simple separable state of order } N   \right\}   \longleftrightarrow $ \\
$\left\{ \mbox{characters of irreducible complex representation of simple separable state group of level }N  \right\} $

{\it Proof \ }
First, reduced expression of simple separable state forms a group, thus reduced expression is a character of one-dimensional complex representation. On the other hand, every irreducible character of finite abelian group is one-dimensional \cite{ALG}. Let $\varphi: G \to \mathbb{C}$ is one-dimensional complex representation, since ${x^{(i)}}^2 = 1, \ x^{(i)} = \pm 1$, then the character of $\varphi$ dose indeed corresponds to a reduced expression of simple separable state.  $\Box$

The orthogonality of irreducible character \cite{GRO} can be used to prove theorems in this paper.  \\

\noindent {\bf Appendix E}

{\bf Theorem S1 \ }
$N$-qubit states are written as the vector representations given in Eq.(3): $(x_{00...01},x_{00...02},...,x_{33...3})$, define $\boldsymbol{u}_{i,j,...,k}^{\pm} := (0,0,..,0,\pm 1,0,...,0)$, only the position corresponding to $x_{i,j,...,k}$ is 1 or -1, while all other positions are 0. We have that: except $i=j=...=k=0$, for all $0 \leq i,j,...,k \leq 3$, $\boldsymbol{u}_{i,j,...,k}^{+}$ and $\boldsymbol{u}_{i,j,...,k}^{-}$ are all fully separable. Transform into density matrix, that is, any $\frac{1}{2^N} \left (
I \pm \sigma_i \otimes \sigma_j \otimes ... \otimes \sigma_k \right )$ is fully separable quantum state, where $\sigma_0 = I$.

{\bf Proof of Theorem S1}

The proof of Theorem S1 in this Appendix based on Appendix D. First, we prove two lemmas.

{\bf Lemma S3 \ }
For $N$-qubit system, $ \left\{ s_1,s_2,...,s_{2^{N}} \right\}$ are all reduced expressions of simple separable states, then
$$ \sum_{i=1}^{2^N} \frac{1}{2^N} s_i = (1,0,0,...,0) .$$

{\it Proof \ }
By Lemma S2 in Appendix D, $ \left\{ s_1,s_2,...,s_{2^{N}} \right\}$ are all characters of irreducible complex representation of simple separable state group. The columns of the character table are orthogonal \cite{GRO}. The first column is $(1,1,1,...,1)^\mathrm{ T } $, thus $(1,1,1,...,1) \cdot (s_1,s_2,s_3,...,s_{2^{N}})^\mathrm{T} = (2^N,0,...,0)$, where $^\mathrm{T}$ is transpose. The lemma is proved. $\Box$ \\

{\bf Lemma S4 \ }
$G$ is simple separable group of level $N$. $$G = \left< x^{(1)}, x^{(2)}, ... , x^{(N)} \ \big| \ {x^{(i)}}^2 = 1, \ x^{(i)} x^{(j)} = x^{(j)} x^{(i)} \ \mbox{for all} \ 1 \leq i,j \leq N \right> $$
where $G$ is a group generated by the generators $\left\{ x^{(1)}, x^{(2)}, ... , x^{(N)} \right\}$ and the equivalence relations are $ {x^{(i)}}^2 = 1, \ x^{(i)} x^{(j)} = x^{(j)} x^{(i)}$ for all $1 \leq i,j \leq N$. And
$$G= \left\{ 1, x^{(1)}, x^{(2)}, ... , x^{(N)} ,x^{(1)}x^{(2)}, x^{(1)}x^{(3)}, ... ,x^{(N-1)}x^{(N)}, x^{(1)}x^{(2)}x^{(3)},...,x^{(1)}x^{(2)}... \ x^{(N)}  \right\} \ ,$$
for any group element $g := x^{(t_1)}x^{(t_2)}... \ x^{(t_m)} \in G$, $1 \leq t_1 < t_2 <...< t_m \leq N$, there is:
$$ G = \left< x^{(1)}, x^{(2)}, ... , x^{(N)} \right> \Big / \sim \; \; = \; \;  \left< g, x^{(1)}, x^{(2)}, ... ,\hat{x^{(t_1)}}, ... , x^{(N)} \right> \Big / \sim $$
where $\hat{x^{(t_1)}}$ denotes the removal of this term, that is, replacing $x^{(t_1)}$ with $x^{(t_1)}x^{(t_2)}... \ x^{(t_m)}$ in the generators set, $\sim$ is the above equivalence relation.

{\it Proof \ }
Due to $g \in G$,
$$ \left< x^{(1)}, x^{(2)}, ... , x^{(N)} \right> \Big / \sim \;  \; \supseteq  \;  \;  \left< g, x^{(1)}, x^{(2)}, ... ,\hat{x^{(t_1)}}, ... , x^{(N)} \right> \Big / \sim $$
and $x^{(t_1)} = g x^{(t_2)}x^{(t_3)}... \ x^{(t_m)} $, then
$$ \left< x^{(1)}, x^{(2)}, ... , x^{(N)} \right> \Big / \sim \;  \; \subseteq   \;  \;  \left< g, x^{(1)}, x^{(2)}, ... ,\hat{x^{(t_1)}}, ... , x^{(N)} \right> \Big / \sim $$
the lemma is proved.  $\Box$  \\

{\bf Lemma S5 \ }
$G$ is simple separable group of level $N$,

$G= \left\{ 1, x^{(1)}, x^{(2)}, ... , x^{(N)} ,x^{(1)}x^{(2)}, x^{(1)}x^{(3)}, ... ,x^{(N-1)}x^{(N)}, x^{(1)}x^{(2)}x^{(3)},...,x^{(1)}x^{(2)}... \ x^{(N)}  \right\}$. For any group element $g = x^{(t_1)}x^{(t_2)}... \ x^{(t_m)} \in G$, $1 \leq t_1 < t_2 <...< t_m \leq N$, by Lemma S4, we have
$$G = \left< g, x^{(1)}, x^{(2)}, ... ,\hat{x^{(t_1)}}, ... , x^{(N)} \ \big| \ {x^{(i)}}^2 = 1, \ x^{(i)} x^{(j)} = x^{(j)} x^{(i)} \ \mbox{for all} \ 1 \leq i,j \leq N \right> ,$$
where $G$ is a finite group generated by the generators $\left\{ g, x^{(1)}, x^{(2)}, ... ,\hat{x^{(t_1)}}, ... , x^{(N)} \right\}$ and the equivalence relations are $ {x^{(i)}}^2 = 1, \ x^{(i)} x^{(j)} = x^{(j)} x^{(i)}$ for all $1 \leq i,j \leq N$. Then we have
$$G \cong \mathbb{Z}/2\mathbb{Z} \times G' \ ,$$
where $G' =  \left< x^{(1)}, x^{(2)}, ... ,\hat{x^{(t_1)}}, ... , x^{(N)} \ \big| \ {x^{(i)}}^2 = 1, \ x^{(i)} x^{(j)} = x^{(j)} x^{(i)} \ \mbox{for all} \ 1 \leq i,j \leq N \right>$.

{\it Proof \ }
It is straightforward to verify that
\begin{align} \notag
G & \cong \left< g \big| {x^{(i)}}^2 = 1, \ x^{(i)} x^{(j)} = x^{(j)} x^{(i)} \right> \times \left< x^{(1)}, ... ,\hat{x^{(t_1)}}, ... , x^{(N)} \ \big| \ {x^{(i)}}^2 = 1, \ x^{(i)} x^{(j)} = x^{(j)} x^{(i)} \right>  \\
& \cong \left< g \big| {x^{(i)}}^2 = 1, \ x^{(i)} x^{(j)} = x^{(j)} x^{(i)} \right> \times G' \ ,  \notag
\end{align}
and $g = x^{(t_1)}x^{(t_2)}... \ x^{(t_m)} $, so $\left< g \big| {x^{(i)}}^2 = 1, \ x^{(i)} x^{(j)} = x^{(j)} x^{(i)} \right> = \left< g \big| g^2=1 \right> \cong \mathbb{Z}/2\mathbb{Z}$. The lemma is proved.  $\Box$  \\

Now we begin to prove Theorem S1. $G$ is simple separable state group of level $N$. Next, we construct $\boldsymbol{u}_{i,j,...,k}^{+}$ and $\boldsymbol{u}_{i,j,...,k}^{-} \ $. $\boldsymbol{u}_{i,j,...,k}$ corresponds to a group element of $G$, for example, $\boldsymbol{u}_{0,0,...,0,a}$ corresponds to $x^{(N)}$, $\boldsymbol{u}_{a,0,b,0,...,0}$ corresponds to $x^{(1)}x^{(3)}$, $\boldsymbol{u}_{a,b,c,d,...,f}$ corresponds to $x^{(1)}x^{(2)}... \ x^{(N)}$ and so on. Recall the definition of reduced expressions in Appendix D, where multiple quantum states can correspond to the same reduced expressions. Therefore, this correspondence is independent of the specific values of $a,b,c,...$. Without losing generality, let $\boldsymbol{u}_{i,j,...,k}$ corresponds to $x^{(t_1)}x^{(t_2)}... \ x^{(t_m)} = g$. In the character table, $s_u \left ( g \right ) = \pm 1$, where $\left\{ s_u \right\}$ are characters of $G$. We divide the characters into two distinct classes: one where $s_u \left ( g \right ) =  1$, denote as $s_u^+$; and another $s_u \left ( g \right ) = -1$, denote as $s_u^-$.

We claim that:
\begin{align}\label{character+}
\sum_u \frac{1}{2^{N-1}} s_u^+ = (1,0,...,0,1,0,...,0) ,\\
\sum_u \frac{1}{2^{N-1}} s_u^- = (1,0,...,0,-1,0,...,0), \label{character-}
\end{align}
where the component $s_u (g) = \pm 1$, all other components is 0.

Before we proceed with the proof, let's examine some examples to build intuition and then continue with the formal proof. Recall Eq.(9), for 2-qubit, when we choose $\boldsymbol{u}_{i,0}^{+}$, it corresponds to group element $x$, then $\left\{ s_u^+ \right\} = \left\{ s_1, s_2 \right\}$, $\left\{ s_u^- \right\} = \left\{ s_3, s_4 \right\}$. $\sum_u \frac{1}{2}s_u^+ = \frac{1}{2}s_1 + \frac{1}{2}s_2 = (1,1,0,0)$, it is a reduced expression of 3 quantum states: $\boldsymbol{u}_{1,0}^{+} = \frac{1}{4} \left ( I + \sigma_1 \otimes I \right ), \boldsymbol{u}_{2,0}^{+} = \frac{1}{4} \left ( I + \sigma_2 \otimes I \right ), \boldsymbol{u}_{3,0}^{+} = \frac{1}{4} \left ( I + \sigma_3 \otimes I \right )$. $\sum_u \frac{1}{2}s_u^- = \frac{1}{2}s_3 + \frac{1}{2}s_4 = (1,-1,0,0)$, it is also a reduced expression of 3 quantum states: $\boldsymbol{u}_{i,0}^{-} = \frac{1}{4} \left ( I - \sigma_i \otimes I \right )$. (Note that, to be precise, $\boldsymbol{u}_{i,0}^{-}$ is a reduced expression of $\frac{1}{4} \left ( I - \sigma_i \otimes I \right )$, and we use the equal sign here to indicate that it does not cause problems.) Obviously, the above 6 quantum states are separable. Specifically, for $\boldsymbol{u}_{1,0}^{+} = \frac{1}{4} \left ( I + \sigma_1 \otimes I \right )$,
$$
\frac{1}{2} \left ( \frac{1}{4} \left ( I + \sigma_1 \otimes I + I \otimes \sigma_i + \sigma_1 \otimes \sigma_i \right )  +  \frac{1}{4} \left ( I + \sigma_1 \otimes I - I \otimes \sigma_i - \sigma_1 \otimes \sigma_i \right ) \right )  = \frac{1}{4} \left ( I + \sigma_1 \otimes I \right )$$
above equation corresponds to $\frac{1}{2} \left ( s_1 + s_2 \right )= (1,1,0,0)$. The provided examples indeed suggest that Theorem S1 holds for the case when $N=2$. Next, for $N=3$ the character table is given in Appendix G, if we want to construct $(1,0,1,0,0,0,0,0)$, we choose $\left\{ s_u^+ \right\} = \{s_1, s_2, s_5, s_6 \}$, they are are all characters which component at position $y$ is $1$ (that is $s_u(y) = 1$). Then $\sum_u \frac{1}{2^{N-1}} s_u^+ = \frac{1}{4} (s_1+s_2+s_5+s_6) = (1,0,1,0,0,0,0,0)$. It is easy to verify that the claim is true when $N = 3$.

Next, we continue with the formal proof of the claim. Define $g = x^{(t_1)}x^{(t_2)}... \ x^{(t_m)}$, proving that $\sum_u \frac{1}{2^{N-1}} s_u^+ = (1,0,...,0,1,0,...,0), \ \sum_u \frac{1}{2^{N-1}} s_u^- = (1,0,...,0,-1,0,...,0)$ is nothing more than showing that component at position $g$ is $1$ or $-1$ and all other components are 0. Obviously, since $s_u \left ( g \right ) = \pm 1$, the component at position $g$ is $\pm 1$. Next prove that all other components are 0. By Lemma S5 we have $G \cong \mathbb{Z}/2\mathbb{Z} \times G' $, and $G' =  \left< x^{(1)}, x^{(2)}, ... ,\hat{x^{(t_1)}}, ... , x^{(N)} \ \big| \ {x^{(i)}}^2 = 1, \ x^{(i)} x^{(j)} = x^{(j)} x^{(i)} \ \mbox{for all} \ 1 \leq i,j \leq N \right>$ is simple separable group of level $N-1$. Denote all characters of $G'$ as $\left\{ s'_i \right\}$, by Lemma S3 we have
$$ \sum_{i=1}^{2^{N-1}} \frac{1}{2^{N-1}} s'_i(h) = 0 , \ \mbox{for all} \ 1 \neq h \in G' \ .$$
For any group element $gh \in G \ , 1 \neq h \in G'$, by the theorem of characters of direct product group (see Proposition 4.5.1. in \cite{GRO}) and Lemma S5, we have that for any character $s_u$ there exists a character $s'_i$ such that $s_u(g h) =  s_u(g) s'_i(h)$, then
$$s_u^+(g h) =  s_u^+(g) s'_i(h) = s'_i(h) \ ,$$
thus we have
$$ \sum_{i=1}^{2^{N-1}} \frac{1}{2^{N-1}} s_u^+(g h) = \sum_{i=1}^{2^{N-1}} \frac{1}{2^{N-1}} s'_i(h) = 0 ,$$
we have now shown that all components of $\sum_{i=1}^{2^{N-1}} \frac{1}{2^{N-1}} s_u^+$ except $g$-position are 0. Exactly the same,
$$ \sum_{i=1}^{2^{N-1}} \frac{1}{2^{N-1}} s_u^-(g h) = \sum_{i=1}^{2^{N-1}} \frac{1}{2^{N-1}} s_u^-(g) s'_i(h) =  \sum_{i=1}^{2^{N-1}} \frac{-1}{2^{N-1}} s'_i(h) = 0 ,$$
thus Eq.(\ref{character+})(\ref{character-}) is proved.

Finally, recall Appendix D and the beginning of this proof, each reduced expression $s_u^\pm$ corresponds to a simple separable state. It is then easy to verify that a convex combination of reduced expressions $s_u^+$ corresponds to a convex combination of simple separable states, and $\sum_{u=1}^{2^{N-1}} \frac{1}{2^{N-1}} s_u^+ $ corresponds to $\boldsymbol{u}_{i,j,...,k}^+$ (according to the correspondence determined by $\boldsymbol{u}_{i,j,...,k}^+$). So $\boldsymbol{u}_{i,j,...,k}^+$ is a convex combination of simple separable states by Eq.(\ref{character+})(\ref{character-}). $\boldsymbol{u}_{i,j,...,k}^+$ and $\boldsymbol{u}_{i,j,...,k}^-$ are separable, Theorem S1  is proved.  $\Box$ \\

\noindent {\bf Appendix F}

{\bf Proof of Theorem 3}

Note that, similar to the previous section, the computation of the 1-norm omits first component $\rho_{0,...,0} = 1$. If $\left\| \rho \right\|_1 \leq 1$, then
$$\rho = \left (  \sum_{i,j,...,k} |\rho_{i,j,...,k}| \cdot Sgn(\rho_{i,j,...,k}) \cdot \boldsymbol{u}_{i,j,...,k}^+ \right ) + \left (
1 - \left\| \rho \right\|_1 \right ) \cdot \boldsymbol{0} ,$$
where ${\boldsymbol{u}}_{i,j,...,k}^{\pm}$ is defined in Theorem S1 in Appendix E, $Sgn(\rho_{i,j,...,k}) \cdot \boldsymbol{u}_{i,j,...,k}^+ = {\boldsymbol{u}}_{i,j,...,k}^{\pm}$, $\boldsymbol{0}$ is zero vector, and $\boldsymbol{0}$ corresponds to density matrix $\frac{1}{2^N} I$, by Theorem S1 in Appendix E, they are all fully separable, thus $\rho$ is fully separable. The separable decomposition of the density operator $\varrho$ corresponding to $\rho$ is given below:
$$\varrho = \left (  \sum_{i,j,...,k} |\rho_{i,j,...,k}| \frac{1}{2^N} \left (
I + Sgn(\rho_{i,j,...,k}) \sigma_i \otimes \sigma_j \otimes ... \otimes \sigma_k \right ) \right ) + \left (
1 - \left\| \rho \right\|_1 \right ) \frac{1}{2^N} I \ ,$$
where $\sigma_0 =I$, $\sigma_i$ is the Pauli matrix. $\Box$ \\

\noindent {\bf Appendix G}

{\bf Proof of Theorem 4}

The proof in this Appendix based on Appendix D. By identifying more separable pure states, stronger sufficient conditions for separability can be obtained. Similar to Eq.(9), all reduced expressions of 3-qubit simple separable states can be expressed as:

\begin{equation*} 
    \bordermatrix{%
      &\overset{\rho_{000}}{1}   &\overset{\rho_{i00}}{x}   &\overset{\rho_{0j0}}{y}    &\overset{\rho_{00k}}{z}    &\overset{\rho_{ij0}}{xy}   &\overset{\rho_{i0k}}{xz}     &\overset{\rho_{0jk}}{yz}    &\overset{\rho_{ijk}}{xyz}      \cr
s_1   & 1   & 1   & 1   & 1   & 1   & 1     & 1    & 1      \cr
s_2   & 1   & 1   & 1   &-1   & 1   &-1     &-1    &-1      \cr
s_3   & 1   & 1   &-1   & 1   &-1   & 1     &-1    &-1      \cr
s_4   & 1   & 1   &-1   &-1   &-1   &-1     & 1    & 1      \cr
s_5   & 1   & -1  & 1   & 1   &-1   &-1     & 1    &-1       \cr
s_6   & 1   & -1  & 1   &-1   &-1   & 1     &-1    & 1       \cr
s_7   & 1   & -1  &-1   & 1   & 1   &-1     &-1    & 1       \cr
s_8   & 1   & -1  &-1   &-1   & 1   & 1     & 1    &-1       \cr
    }
\end{equation*}
$\frac{1}{2} \left ( s_1+s_8 \right ) = (1,0,0,0,1,1,1,0) \ ; \ \ $
$\frac{1}{2} \left ( s_2+s_7 \right ) = (1,0,0,0,1,-1,-1,0) \ ; \ \ $ \\
$\frac{1}{2} \left ( s_3+s_6 \right ) = (1,0,0,0,-1,1,-1,0) \ ; \ \ $
$\frac{1}{2} \left ( s_4+s_5 \right ) = (1,0,0,0,-1,-1,1,0) \ $, each of the above expressions correspond to 27 separable quantum states, for example,\\
$(1,0,0,0,1,-1,-1,0)$ correspond to density matrices
\begin{align}
\left\{ \frac{1}{8} \left ( I_8 + \sigma_i \otimes \sigma_j \otimes I - \sigma_i \otimes I \otimes \sigma_k - I \otimes \sigma_j \otimes \sigma_k  \right ) \mid 1 \leq i,j,k \leq 3 \right\}.
\end{align}
Using these separable states, a stronger criterion can be derived as follows. In this proof, density matrix of quantum states is used instead of the vector representation. Since the indices $i,j,k$ are symmetric, without loss of generality, let $|\rho_{0,j,k}| \geq  |\rho_{i,0,k}| \geq |\rho_{i,j,0}|$. If $|\rho_{i,j,0}|=0$, then Theorem 4 degenerates to Theorem 3. Thus it is sufficient to prove the $|\rho_{0,j,k}| \geq  |\rho_{i,0,k}| \geq |\rho_{i,j,0}| >0$ case. By Eq.(2), $\varrho = \frac{1}{8} \left ( I_8 + \sum_{r,s,t} \rho_{r,s,t} \cdot \sigma_{r,s,t} \right )$, where $\sigma_{r,s,t} = \sigma_r \otimes \sigma_s \otimes \sigma_t, \ \sigma_0 = I, \ 0 \leq r,s,t \leq 3 \ \mbox{are not all zero}$. Isolate the $i,j,k$ in the summation,
\begin{align}
\varrho =  \frac{1}{8} I_8 + \frac{1}{8} \left ( \rho_{0,j,k} \cdot \sigma_{0,j,k} + \rho_{i,0,k} \cdot\sigma_{i,0,k} + \rho_{i,j,0} \cdot \sigma_{i,j,0} \right ) + \frac{1}{8}{\sum_{r,s,t}}' \rho_{r,s,t} \cdot \sigma_{r,s,t}
\end{align}
where
\begin{align} \notag
& \; \; \; \; \ \rho_{0,j,k} \cdot \sigma_{0,j,k} + \rho_{i,0,k} \cdot\sigma_{i,0,k} + \rho_{i,j,0} \cdot \sigma_{i,j,0} \notag \\
& =  |\rho_{0,j,k}| Sgn(\rho_{0,j,k}) \cdot \sigma_{0,j,k} + |\rho_{i,0,k}| Sgn(\rho_{i,0,k}) \cdot \sigma_{i,0,k} + |\rho_{i,j,0}| Sgn(\rho_{i,j,0}) \cdot \sigma_{i,j,0} \notag \\
& =  |\rho_{i,0,k}| \left ( Sgn(\rho_{0,j,k}) \sigma_{0,j,k}
+ Sgn(\rho_{i,0,k}) \sigma_{i,0,k} + Sgn(\rho_{i,j,0}) \sigma_{i,j,0}  \right ) \notag \\
& \; \; \; \; \ + \left ( |\rho_{0,j,k}| - |\rho_{i,0,k}| \right ) Sgn(\rho_{0,j,k}) \sigma_{0,j,k} + \left ( |\rho_{i,0,k}| - |\rho_{i,j,0}| \right ) \left ( - Sgn(\rho_{i,j,0}) \sigma_{i,j,0} \right ) , \notag
\end{align}
as with the separable states in Eq.(12),
$$ \frac{1}{8} \left ( I_8 + Sgn(\rho_{0,j,k}) \sigma_{0,j,k} + Sgn(\rho_{i,0,k}) \sigma_{i,0,k} + Sgn(\rho_{i,j,0}) \sigma_{i,j,0} \right )$$
is separable. By Theorem S1 in Appendix E, $\frac{1}{8} \left ( I_8 + Sgn(\rho_{i,0,k}) \sigma_{i,0,k} \right )$, $ \frac{1}{8} \left ( I_8 - Sgn(\rho_{i,j,0})  \sigma_{i,j,0} \right ) \ $ and $ \frac{1}{8} \left ( I_8 - Sgn(\rho_{r,s,t})  \sigma_{r,s,t} \right ) \ $ are separable. Therefore if $\left\| \rho \right\|_1 - 2min\left\{ |\rho_{0,j,k}|, |\rho_{i,0,k}|, |\rho_{i,j,0}| \right\} \leq 1 $, then $\rho$ can be written as a convex combination of separable states by Eq.(13):
\begin{align} \notag
\varrho = & |\rho_{i,0,k}| \frac{1}{8} \left ( I_8 + Sgn(\rho_{0,j,k}) \sigma_{0,j,k}
+ Sgn(\rho_{i,0,k}) \sigma_{i,0,k} + Sgn(\rho_{i,j,0}) \sigma_{i,j,0}  \right )  \\
+ & \left ( |\rho_{0,j,k}| - |\rho_{i,0,k}| \right ) \frac{1}{8} \left ( I_8 + Sgn(\rho_{0,j,k}) \sigma_{0,j,k} \right )  \notag  \\
+ & \left ( |\rho_{i,0,k}| - |\rho_{i,j,0}| \right ) \frac{1}{8} \left ( I_8 - Sgn(\rho_{i,j,0})  \sigma_{i,j,0} \right ) \notag  \\
+ &  {\sum_{r,s,t}}' |\rho_{r,s,t}| \frac{1}{8} \left ( I_8 + Sgn(\rho_{r,s,t}) \sigma_{r,s,t} \right ) + \left ( 1 - \left\| \rho \right\|_1 + 2|\rho_{i,j,0}| \right ) \frac{1}{8}I_8 . \notag
\end{align}
Finally, it is proven that $\rho$ is separable. $\Box$ \\

{\bf Proof of Theorem 5}

In exactly the same way as in the proof above, without loss of generality, let $|\rho_{0,j_t,k_t}| \geq  |\rho_{i_t,0,k_t}| \geq |\rho_{i_t,j_t,0}| >0$ for all $1 \leq t \leq 3$, if $ \left\| \rho \right\|_1 \leq 1 + 2 |\rho_{i_1,j_1,0}|  + 2 |\rho_{i_2,j_2,0}| +  2|\rho_{i_3,j_3,0}| , \notag $
then $\varrho$ can be expressed as a convex combination of separable states:
\begin{align} \notag
\varrho = & |\rho_{i_1,0,k_1}| \frac{1}{8} \left ( I_8 + Sgn(\rho_{0,j_1,k_1}) \sigma_{0,j_1,k_1}
+ Sgn(\rho_{i_1,0,k_1}) \sigma_{i_1,0,k_1} + Sgn(\rho_{i_1,j_1,0}) \sigma_{i_1,j_1,0}  \right )  \notag \\
+ & \left ( |\rho_{0,j_1,k_1}| - |\rho_{i_1,0,k_1}| \right ) \frac{1}{8} \left ( I_8 + Sgn(\rho_{0,j_1,k_1}) \sigma_{0,j_1,k_1} \right ) \notag  \\
+ & \left ( |\rho_{i_1,0,k_1}| - |\rho_{i_1,j_1,0}| \right ) \frac{1}{8} \left ( I_8 - Sgn(\rho_{i_1,j_1,0})  \sigma_{i_1,j_1,0} \right ) \notag  \\
+ & |\rho_{i_2,0,k_2}| \frac{1}{8} \left ( I_8 + Sgn(\rho_{0,j_2,k_2}) \sigma_{0,j_2,k_2}
+ Sgn(\rho_{i_2,0,k_2}) \sigma_{i_2,0,k_2} + Sgn(\rho_{i_2,j_2,0}) \sigma_{i_2,j_2,0}  \right )  \notag \\
+ & \left ( |\rho_{0,j_2,k_2}| - |\rho_{i_2,0,k_2}| \right ) \frac{1}{8} \left ( I_8 + Sgn(\rho_{0,j_2,k_2}) \sigma_{0,j_2,k_2} \right ) \notag \\
+ & \left ( |\rho_{i_2,0,k_2}| - |\rho_{i_2,j_2,0}| \right ) \frac{1}{8} \left ( I_8 - Sgn(\rho_{i_2,j_2,0})  \sigma_{i_2,j_2,0} \right ) \notag  \\
+ & |\rho_{i_3,0,k_3}| \frac{1}{8} \left ( I_8 + Sgn(\rho_{0,j_3,k_3}) \sigma_{0,j_3,k_3}
+ Sgn(\rho_{i_3,0,k_3}) \sigma_{i_3,0,k_3} + Sgn(\rho_{i_3,j_3,0}) \sigma_{i_3,j_3,0}  \right )  \notag \\
+ & \left ( |\rho_{0,j_3,k_3}| - |\rho_{i_3,0,k_3}| \right ) \frac{1}{8} \left ( I_8 + Sgn(\rho_{0,j_3,k_3}) \sigma_{0,j_3,k_3} \right )  \notag \\
+ & \left ( |\rho_{i_3,0,k_3}| - |\rho_{i_3,j_3,0}| \right ) \frac{1}{8} \left ( I_8 - Sgn(\rho_{i_3,j_3,0})  \sigma_{i_3,j_3,0} \right ) \notag  \notag \\
+ &  \sum_{r,s,t} |\rho_{r,s,t}| \frac{1}{8} \left ( I_8 + Sgn(\rho_{r,s,t}) \sigma_{r,s,t} \right ) + \left ( 1 - \left\| \rho \right\|_1 + 2 |\rho_{i_1,j_1,0}|  + 2 |\rho_{i_2,j_2,0}| +  2 |\rho_{i_3,j_3,0}| \right ) \frac{1}{8}I_8 . \notag
\end{align} \\

\noindent {\bf Appendix H}

{\bf Proof of Theorem 6}

The proof in this Appendix based on Appendix D. All reduced expressions of 4-qubit simple separable states can be expressed as:

\begin{adjustbox}{scale=0.858}

$$
    \bordermatrix{%
      &\overset{\rho_{0000}}{1} &\overset{\rho_{i000}}{w} &\overset{\rho_{0j00}}{x}  &\overset{\rho_{00k0}}{y}  &\overset{\rho_{000l}}{z}   &\overset{\rho_{ij00}}{wx} &\overset{\rho_{i0k0}}{wy} &\overset{\rho_{i00l}}{wz} &\overset{\rho_{0jk0}}{xy}  &\overset{\rho_{0j0l}}{xz} &\overset{\rho_{00kl}}{yz} &\overset{\rho_{ijk0}}{wxy} &\overset{\rho_{ij0l}}{wxz} &\overset{\rho_{i0kl}}{wyz} &\overset{\rho_{0jkl}}{xyz} &\overset{\rho_{ijkl}}{wxyz} \cr
s_1   & 1  & 1 & 1   & 1   & 1  &1   & 1  & 1 & 1   & 1  & 1 & 1  & 1  & 1 & 1   & 1 \cr
s_2   & 1  & 1 & 1   & 1   &-1  &1   & 1  &-1 & 1   &-1  &-1 & 1  &-1  &-1 &-1   &-1 \cr
s_3   & 1  & 1 & 1   &-1   & 1  &1   &-1  & 1 &-1   & 1  &-1 &-1  & 1  &-1 &-1   &-1 \cr
s_4   & 1  & 1 & 1   &-1   &-1  &1   &-1  &-1 &-1   &-1  & 1 &-1  &-1  & 1 & 1   & 1 \cr
s_5   & 1  & 1 & -1  & 1   & 1  &-1  & 1  & 1 &-1   &-1  & 1 &-1  &-1  & 1 &-1   &-1 \cr
s_6   & 1  & 1 & -1  & 1   &-1  &-1  & 1  &-1 &-1   & 1  &-1 &-1  & 1  &-1 & 1   & 1 \cr
s_7   & 1  & 1 & -1  &-1   & 1  &-1  &-1  & 1 & 1   &-1  &-1 & 1  &-1  &-1 & 1   & 1 \cr
s_8   & 1  & 1 & -1  &-1   &-1  &-1  &-1  &-1 & 1   & 1  & 1 & 1  & 1  & 1 &-1   &-1 \cr
s_9   & 1  &-1 & 1   & 1   & 1  &-1  &-1  &-1 & 1   & 1  & 1 &-1  &-1  &-1 & 1   &-1 \cr
s_{10}& 1  &-1 & 1   & 1   &-1  &-1  &-1  & 1 & 1   &-1  &-1 &-1  & 1  & 1 &-1   & 1 \cr
s_{11}& 1  &-1 & 1   &-1   & 1  &-1  & 1  &-1 &-1   & 1  &-1 & 1  &-1  & 1 &-1   & 1 \cr
s_{12}& 1  &-1 & 1   &-1   &-1  &-1  & 1  & 1 &-1   &-1  & 1 & 1  & 1  &-1 & 1   &-1 \cr
s_{13}& 1  &-1 & -1  & 1   & 1  & 1  &-1  &-1 &-1   &-1  & 1 & 1  & 1  &-1 &-1   & 1 \cr
s_{14}& 1  &-1 & -1  & 1   &-1  & 1  &-1  & 1 &-1   & 1  &-1 & 1  &-1  & 1 & 1   &-1 \cr
s_{15}& 1  &-1 & -1  &-1   & 1  & 1  & 1  &-1 & 1   &-1  &-1 &-1  & 1  & 1 & 1   &-1 \cr
s_{16}& 1  &-1 & -1  &-1   &-1  & 1  & 1  & 1 & 1   & 1  & 1 &-1  &-1  &-1 &-1   & 1 \cr
    }
$$

\end{adjustbox}

\begin{align}
& \frac{1}{2}  \left ( s_1 + s_{16} \right ) = (1,0,0,0,0,1,1,1,1,1,1,0,0,0,0,1);  \notag\\
& \frac{1}{2}  \left ( s_2 + s_{15} \right ) = (1,0,0,0,0,1,1,-1,1,-1,-1,0,0,0,0,-1); \notag\\
& \frac{1}{2}  \left ( s_3 + s_{14} \right ) = (1,0,0,0,0,1,-1,1,-1,1,-1,0,0,0,0,-1); \notag\\
& \frac{1}{2}  \left ( s_4 + s_{13} \right ) = (1,0,0,0,0,1,-1,-1,-1,-1,1,0,0,0,0,1); \notag\\
& \frac{1}{2}  \left ( s_5 + s_{12} \right ) = (1,0,0,0,0,-1,1,1,-1,-1,1,0,0,0,0,-1); \notag\\
& \frac{1}{2}  \left ( s_6 + s_{11} \right ) = (1,0,0,0,0,-1,1,-1,-1,1,-1,0,0,0,0,1); \notag\\
& \frac{1}{2}  \left ( s_7 + s_{10} \right ) = (1,0,0,0,0,-1,-1,1,1,-1,-1,0,0,0,0,1); \notag\\
& \frac{1}{2}  \left ( s_8 + s_{9} \right ) = (1,0,0,0,0,-1,-1,-1,1,1,1,0,0,0,0,-1).
\end{align}

Each of the above reduced expression corresponds to $3^4$ separable quantum states, just like in Eq.(12). Completely analogous to Theorem 4, we have the following proof:

To include the case $\rho_{a,b,c,d} = 0$, define $\widetilde{Sgn}(x) =1$, if $x \geq 0$; $\widetilde{Sgn}(x) = -1$, if $x < 0$. The conditions required in the Theorem 6 guarantee that the sign correspond to a reduced expression in Eq.(14). Without loss of generality, let
$$ |\rho_{i,j,0,0}| \geq  |\rho_{i,0,k,0}| \geq  |\rho_{i,0,0,l}| \geq
 |\rho_{0,j,k,0}| \geq |\rho_{0,j,0,l}| \geq  \ |\rho_{0,0,k,l}| \geq
 |\rho_{i,j,k,l}|$$
define $\sigma_{a,b,c,d} := \sigma_a \otimes \sigma_b \otimes \sigma_c \otimes \sigma_d$, where $\sigma_0 = I, \ \sigma_i$ is Pauli matrix. Completely analogous to Eq.(13), isolate the $i,j,k,l$ in the summation,
\begin{align}
16\varrho = &  I +  \rho_{i,j,0,0} \cdot \sigma_{i,j,0,0} + \rho_{i,0,k,0} \cdot \sigma_{i,0,k,0} + \rho_{i,0,0,l} \cdot \sigma_{i,0,0,l} + \rho_{0,j,k,0} \cdot \sigma_{0,j,k,0} + \rho_{0,j,0,l} \cdot \sigma_{0,j,0,l}  \notag\\
& + \rho_{0,0,k,l} \cdot \sigma_{0,0,k,l} + \rho_{i,j,k,l} \cdot \sigma_{i,j,k,l} + \sum_{r,s,t,q} \rho_{r,s,t,q} \cdot \sigma_{r,s,t,q} \notag \\
= & |\rho_{0,j,k,0}|  \left ( I + \varrho_s \right ) + \left ( |\rho_{i,j,0,0}| - |\rho_{0,j,k,0}| \right )  \left ( I + \widetilde{Sgn}(\rho_{i,j,0,0}) \ \sigma_{i,j,0,0} \right )   \notag \\
& + \left ( |\rho_{i,0,k,0}| - |\rho_{0,j,k,0}| \right )  \left ( I + \widetilde{Sgn}(\rho_{i,0,k,0}) \ \sigma_{i,0,k,0} \right )   \notag \\
& + \left ( |\rho_{i,0,0,l}| - |\rho_{0,j,k,0}| \right )  \left ( I + \widetilde{Sgn}(\rho_{i,0,0,l}) \ \sigma_{i,0,0,l} \right )   \notag \\
&+  \left ( |\rho_{0,j,k,0}| - |\rho_{0,j,0,l}| \right )  \left ( I - \widetilde{Sgn}(\rho_{0,j,0,l}) \ \sigma_{0,j,0,l} \right )   \notag \\
&+  \left ( |\rho_{0,j,k,0}| - |\rho_{0,0,k,l}| \right )  \left ( I - \widetilde{Sgn}(\rho_{0,0,k,l}) \ \sigma_{0,0,k,l} \right )   \notag \\
&+  \left ( |\rho_{0,j,k,0}| - |\rho_{i,j,k,l}| \right )  \left ( I - \widetilde{Sgn}(\rho_{i,j,k,l}) \ \sigma_{i,j,k,l} \right )   \notag \\
&+ \sum_{r,s,t,q} |\rho_{r,s,t,q}| \left ( I + \widetilde{Sgn}(\rho_{r,s,t,q}) \sigma_{r,s,t,q} \right)          \notag \\
&+ \left ( 1 - \left\| \rho \right\|_1 + 2 \left( |\rho_{0,j,0,l}| + |\rho_{0,0,k,l}| + |\rho_{i,j,k,l}|  \right) \ \right ) I
\end{align}
where $\varrho_s := \widetilde{Sgn}(\rho_{i,j,0,0}) \sigma_{i,j,0,0}
+ \widetilde{Sgn}(\rho_{i,0,k,0}) \sigma_{i,0,k,0} + \widetilde{Sgn}(\rho_{i,0,0,l}) \sigma_{i,0,0,l} + \widetilde{Sgn}(\rho_{0,j,k,0}) \sigma_{0,j,k,0} + \widetilde{Sgn}(\rho_{0,j,0,l}) \sigma_{0,j,0,l} + \widetilde{Sgn}(\rho_{0,0,k,l}) \sigma_{0,0,k,l} + \widetilde{Sgn}(\rho_{i,j,k,l}) \sigma_{i,j,k,l} $.

By Theorem S1 in Appendix E and Eq.(14), in the equation above, $\frac{1}{16} \left ( I + \varrho_s \right ) $, \\
$\frac{1}{16} \left ( I \pm \widetilde{Sgn}(\rho_{r,s,t,q}) \sigma_{r,s,t,q}  \right ) $, $\frac{1}{16} I$ are all separable states. So when
$$ 1 - \left\| \rho \right\|_1 + 2 \left( |\rho_{0,j,0,l}| + |\rho_{0,0,k,l}| + |\rho_{i,j,k,l}|  \right) \geq 0 \ , $$
Eq.(15) is a convex combination of separable states, that is, $\rho$ is separable.  $\Box$ \\

\noindent {\bf Appendix I}

For arbitrary $N$-partite quantum system $\mathcal{H}=\mathbb{C}^{n_1} \otimes \mathbb{C}^{n_2} \otimes ... \otimes\mathbb{C}^{n_N}$, changing the normalization of the basis in Eq.(2) with
\begin{align}\
\{ {\lambda '}_{i_1}^{(1)} \otimes {\lambda'}_{i_2}^{(2)} \otimes ... \otimes {\lambda'}_{i_N}^{(N)} \mid 0\leq i_k \leq {n_k}^2-1 \ \mbox{for all} \ 1 \leq k \leq N \} ,
\end{align}
where ${\lambda'}_{0}^{(k)} := \sqrt{n_k-1} I_{n_k}, \; {\lambda'}_{j}^{(k)} := \sqrt{\frac{n_k(n_k-1)}{2}} \lambda_{j}^{(k)} = \widetilde{\lambda}_{j}^{(k)}$, now $\forall i. \ Tr \left ( \left ( {\lambda'}_{i}^{(k)} \right )^2 \right ) = n_k(n_k - 1)$. Using this set of basis, denote $\ n=\prod_{k} n_k, \ {\lambda'}_{i_1,i_2,...,i_N} = {\lambda '}_{i_1}^{(1)} \otimes ... \otimes {\lambda'}_{i_N}^{(N)} $, new vector representation is obtained:
\begin{align}
\varrho = \frac{1}{n}\left ( I_n + \sum_{i_1,i_2...,i_N} \rho_{i_1,i_2,...,i_N} \cdot {\lambda'}_{i_1,i_2,...,i_N} \right) ,
\end{align}
where $\rho_{i_1,i_2,...,i_N}  = n \prod_{k=1}^{N} \frac{1}{Tr \left ( ({\lambda'}_{i_k}^{(k)}) ^2 \right) } \cdot Tr(\varrho \  {\lambda'}_{i_1,i_2,...,i_N} ) = \frac{Tr(\varrho \  {\lambda'}_{i_1,i_2,...,i_N} )}{\prod_{k=1}^{N} ( n_k - 1 )} $. And any pure state $\rho$ satisfies ${ \left\| \rho \right\|_2 }^2 = \frac{( \prod_k n_k ) -1}{\prod_k ( n_k-1 ) }$, detailed proof is given in the next lemma. \\

{\bf Lemma S6 \ }
For arbitrary multipartite quantum system $\mathcal{H}=\mathbb{C}^{n_1} \otimes \mathbb{C}^{n_2} \otimes ... \otimes\mathbb{C}^{n_N}$, by the new extended Bloch representation in Eq.(17),
\begin{align}
Tr(\varrho^2) = \frac{1}{\prod_k n_k}  \left (  1+ \prod_k ( n_k-1 ) \ { \left\| \rho \right\|_2 }^2 \right ) ,
\end{align}
where $\varrho$ denote a density matrix, $\rho$ denote it's vector representation given in Eq.(17). Then any pure state $\varrho$ satisfies ${ \left\| \rho \right\|_2 }^2 = \frac{( \prod_k n_k ) -1}{\prod_k ( n_k-1 ) }$.

{\it Proof \ }
Now $Tr \left ( { {\lambda'}_{i_1,i_2,...,i_N} }^2 \right ) = \prod_k Tr\left ( { {\lambda'}_k }^2 \right ) =  \prod_k \left ( n_k -1 \right ) n_k$, by Eq.(17) we have
\begin{align} \notag
Tr(\varrho ^2) & = \frac{1}{\prod_k {n_k}^2} \left (  \prod_k n_k + Tr \left ( \sum_{i_1,...,i_{N}} {\rho_{i_1,...,i_{N}}}^2  \cdot { {\lambda'}_{i_1,i_2,...,i_N} }^2 \right ) \right ) \notag \\
& = \frac{1}{\prod_k {n_k}^2} \left (  \prod_k n_k + \sum_{i_1,...,i_{N}} {\rho_{i_1,...,i_{N}}}^2  \cdot Tr \left ( { {\lambda'}_{i_1,i_2,...,i_N} }^2  \right )  \right ) \notag \\
& = \frac{1}{\prod_k {n_k}^2} \left ( \prod_k n_k + { \left\| \rho \right\|_2 }^2 \prod_k \left ( n_k -1 \right ) n_k \right ) , \notag
\end{align}
when $\varrho$ is pure state, then we have ${ \left\| \rho \right\|_2 }^2 = \frac{( \prod_k n_k ) -1}{\prod_k ( n_k-1 ) }$. $\Box$ \\

Despite the fact that there are more parameters, the 2-norm of vector representation and purity of quantum state are equivalent and seem to be a more reasonable choice of basis. \\

\noindent {\bf Appendix J}

Theorem 4 and Theorem 6 can be generalized to any $N$-qubit system. Specifically, we can construct separable states as in Eq.(12):
\begin{align} \notag
& \frac{I^{\otimes N}}{2^{N}} + \frac{1}{2^{N}}\left ( \sigma_{i_1} \otimes \sigma_{i_2} \otimes I^{\otimes N-2}   +  \sigma_{i_1} \otimes I \otimes \sigma_{i_3} \otimes I^{\otimes N-3}  + ... + I^{\otimes N-2} \otimes \sigma_{i_{N-1}} \otimes \sigma_{i_{N}}  \right )  \\
& + \frac{1}{2^{N}}\left ( \sigma_{i_1} \otimes \sigma_{i_2} \otimes \sigma_{i_3} \otimes \sigma_{i_4} \otimes I^{\otimes N-4}  + ... + I^{\otimes N-4} \otimes \sigma_{i_{N-3}} \otimes \sigma_{i_{N-2}} \otimes \sigma_{i_{N-1}} \otimes \sigma_{i_{N}}  \right ) \notag \\
& + ...... \notag \\
& + \frac{1}{2^{N}}\left (  \sigma_{i_1} \otimes ... \otimes \sigma_{i_{N-1}} \otimes I + \sigma_{i_1} \otimes ... \otimes \sigma_{i_{N-2}} \otimes I \otimes \sigma_{i_{N}}  + ... + I \otimes \sigma_{i_2} \otimes ... \otimes \sigma_{i_{N}}  \right ) ,
\end{align}
where $1 \leq i_k \leq 3$.

In above equation, the sum is all terms in which the $\sigma_{i_k}$ appears even number of times. (So Eq.(19) is for $N$ being odd, and if $N$ is even, the last term of the summation will have no positions that are $I$). It is corresponds to $\frac{1}{2}s_1 + \frac{1}{2}s_{2^N}$, where $s_1, \ s_{2^N}$ are characters of $N$-level simple separable state group, see Appendix D. In Examples 4 and 5, it is verified that Theorems 4 and 6 are sufficient and necessary criteria for noisy GHZ states. Furthermore, since $N$-qubit GHZ pure state is $\varrho_{GHZ,N} = \frac{1}{2^{N+1}}  \left ( (I + \sigma_3 )^{\otimes N} + (I - \sigma_3 )^{\otimes N} +
(\sigma_1 + i \sigma_2 )^{\otimes N} + (\sigma_1 - i\sigma_2 )^{\otimes N} \right )$ \cite{GHZ}, by Theorem S1 in Appendix E, let $i_1 = i_2 =...= i_N = 3$ in Eq.(19), we can sufficiently and necessarily detect entanglement of any $N$-qubit noisy GHZ state. Specifically, we can construct a separable decomposition of any noisy GHZ state just like Eq.(15) if and only if the noisy GHZ state is separable. In other words, the criterion obtained by generalizing Theorem 4 and Theorem 6 is a sufficient and necessary criterion for $N$-qubit noisy GHZ states.   \\

\noindent {\bf Appendix K}

{\bf Proof of Theorem 9}

We construct the state with the largest $p$-norm under separability restriction. Discussed in two scenarios.

When $1\leq p \leq 2$, consider qubit state $\varrho_1 = \frac{1}{2} \left ( I + \frac{1}{\sqrt{3}}(\sigma_1 + \sigma_2 + \sigma_3  ) \right )$, then ${\varrho_1}^{\otimes N}$ is separable $N$-qubit pure state. $\rho_1$ is the vector representation of ${\varrho_1}^{\otimes N}$. We claim that ${\varrho_1}^{\otimes N}$ is the separable state with the largest $p$-norm. First, by the convexity of norm, the mixing of separable states does not increase norm, so we only need to consider separable pure states. Recall Lemma 1, by Eq.(4), $\left\| \rho \right\|_p$ takes its maximum value when $\left\| \boldsymbol{r}_1 \right\|_p = \left\| \boldsymbol{r}_2 \right\|_p = ... = \left\| \boldsymbol{r}_N \right\|_p$ takes its maximum value. Since they are pure state, $\left\| \boldsymbol{r}_1 \right\|_2 = \left\| \boldsymbol{r}_2 \right\|_2 = ... = \left\| \boldsymbol{r}_N \right\|_2 = 1$, thus $\left\| \rho \right\|_p$ takes its maximum value when:
\begin{align}
\begin{cases} \boldsymbol{r}_1 =...= \boldsymbol{r}_N = \left ( \frac{1}{\sqrt{3}}, \frac{1}{\sqrt{3}}, \frac{1}{\sqrt{3}} \right ) , & \text{if} \ 1 \leq p \leq 2, \\
\boldsymbol{r}_1 =...= \boldsymbol{r}_N = \left ( 1, 0, 0 \right ), & \text{if} \ p>2.  \end{cases}
\end{align}
So ${\varrho_1}^{\otimes N}$ is indeed the separable state with the largest $p$-norm.

When $p > 2$, consider qubit state $\varrho_0 = \frac{1}{2} \left ( I + \sigma_1 \right )$, then ${\varrho_0}^{\otimes N}$ is separable $N$-qubit pure state. $\rho_0$ is the vector representation of ${\varrho_0}^{\otimes N}$. Similarly, by Eq.(20), ${\varrho_0}^{\otimes N}$ is the separable state with the largest $p$-norm. Finally, the minimum $l_p$ balls containing all separable states are given by $\left\| \rho_1 \right\|_p$ and $\left\| \rho_0 \right\|_p$, respectively:
$$R_E(l_p) = \begin{cases} \left ( \left( 3^{1 - \frac{p}{2}}  +1 \right )^N - 1 \right )^{\frac{1}{p}}, & \text{if} \ 1 \leq p \leq 2, \\ \left ( 2^N - 1\right )^{\frac{1}{p}} , & \text{if} \ p > 2. \end{cases}$$

Note that ${\varrho_1}^{\otimes N}$ and ${\varrho_0}^{\otimes N}$ also reach the upper bound of Theorem 1. When $p\geq 2$, one can prove that the minimum $l_p$ ball contains all $N$-qubit states by Theorem S6 in Appendix I. We conjecture that the minimum $l_p$ ball is the smallest $l_p$ ball that contains all $N$-qubit states for $1 \leq p < 2$. For pure states, i.e., $\left\| \rho \right\|_2 = \sqrt{2^N-1}$, the more evenly distributed the components of the vector representation are, the larger their $p$-norm. In the semi-positive definite restriction, we did not find a $3$-qubit state with a larger $p$-norm than ${\varrho_1}^{\otimes 3}$.  \\

\noindent {\bf Appendix L}

The calculations of the criteria we present are all very simple  and seem difficult only because of the complexity of the notation. For the $N$-qubit example below, Pauli matrix is the basis of Bloch representation, lower indicator $0$ corresponds to $I$, lower indicator 1 corresponds to $\sigma_1$, lower indicator 2 corresponds to $\sigma_2$, and lower indicator 3 corresponds to $\sigma_3$.

{\bf Example 6 }
Now consider a $2 \otimes 4$ bound entangled state in Ref.\cite{S}'s Eq.(20):
$\varrho_{a,\alpha} = \alpha \varrho_A + \frac{1-\alpha}{8}I_8$, where $\varrho_A = \frac{7a}{7a+1} \varrho_{ent} + \frac{1}{7a+1} \ket{\phi} \bra{\phi}, \ a \in \left [ 0,1 \right ]$, $\ket{\phi} = \ket{1} \otimes \left ( \sqrt{\frac{1+a}{2}} \ket{0} + \sqrt{\frac{1-a}{2}} \ket{2} \right )$, $\varrho_{ent} = \frac{2}{7}\sum_{i=1}^3 \ket{\psi_i} \bra{\psi_i} + \frac{1}{7} \ket{03} \bra{03}$, $\ket{\psi_i} = \frac{1}{\sqrt{2}} \left ( \ket{0} \ket{i-1} + \ket{1} \ket{i}  \right ), i=1,2,3$. By 3-qubit Bloch representation, the vector representation of $\varrho_{a}$ is: $\rho_{003} = \frac{1-a}{1+7a}, \ \rho_{300} = \rho_{303} = -\frac{1-a}{1+7a}$, $\rho_{010} = \rho_{013} = \frac{\sqrt{1-a^2}}{1+7a}, \ \rho_{202} = -\frac{4a}{1+7a}, \ \rho_{221} = -\frac{2a}{1+7a}$, $\rho_{212} = \rho_{122} = \rho_{111} = \frac{2a}{1+7a} $, $\rho_{101} = \frac{4a}{1+7a} $, $\rho_{310} = \rho_{313} = -\frac{\sqrt{1-a^2}}{1+7a}$. Let $i=3, \ j= 2, \ k=3$ in Theorem 4, then $\left\| \rho_a \right\|_1 - 2 min\left\{ |\rho_{013}|, \ |\rho_{303}|, \ |\rho_{310}| \right\} = \frac{1}{1+7a} \left ( 1+15a+4 \sqrt{1-a^2} \right )$, denote as $f(a)$. $f(a)$ takes its maximum value at $a = 0$, with maximum value $5$. By Theorem 4, when $\alpha \leq \frac{1}{5}$, $\varrho_{a,\alpha}$ is fully separable for all $0 \leq a \leq 1$.
\\

\noindent

\bigskip
\noindent{\bf Acknowledgments:}
This work is supported by the National Natural Science Foundation of China (NSFC) under Grants 12075159 and 12171044, the specific research fund of the Innovation Platform for Academicians of Hainan Province. We are also grateful to the anonymous referees for their suggestions for revisions.

\noindent{\bf Conflicts of Interest:}
The authors declare no other conflict of interest.

\noindent{\bf Data Availability Statement:}
This is no data generated in research.





\end{document}